\documentstyle[fleqn,twoside]{article} 
\begin{document}

\hfill \vbox{\halign{& # \hfill            \cr
		     & OHSTPY-HEP-T-96-001 \cr
		     & MADPH-96-927        \cr
		     & UCD-96-2            \cr
		     & hep-ph/9602374      \cr
		     & February 1996       \cr}}

\pagestyle{myheadings}
\markboth{\rm BRAATEN, FLEMING \& YUAN}
	{\rm QUARKONIUM PRODUCTION}

\medskip
\noindent
{\huge PRODUCTION OF}

\bigskip
\noindent
{\huge HEAVY QUARKONIUM}

\medskip
\noindent
{\huge IN HIGH ENERGY COLLIDERS}

\bigskip
\noindent
{\large {\it Eric Braaten}}

\medskip
\noindent
Department of Physics, Ohio State University, 
Columbus, OH 43210, USA

\medskip
\noindent
{\large {\it Sean Fleming}}

\medskip
\noindent 
Department of Physics, University of Wisconsin, 
Madison, WI 53706, USA

\medskip
\noindent
{\large {\it Tzu Chiang Yuan}}

\medskip
\noindent 
Davis Institute for High Energy Physics,\\
University of California, Davis, CA 95616, USA

\bigskip
\noindent
{\sc key words}:\quad  
bottomonium, charmonium, color-singlet, color-octet,\\
factorization, fragmentation

\bigskip

\tableofcontents
\bigskip
\hrule
\bigskip
\begin{abstract}

Recent data from the Tevatron has revealed that the 
production rate of prompt charmonium at large transverse momentum
is  orders of magnitude larger than the 
best theoretical predictions of a few years ago.
These surprising results can be understood by taking into account
two recent developments that have revolutionized 
the theoretical description of heavy quarkonium production.  
The first is the realization that fragmentation 
must dominate at large transverse momentum, which implies that most
charmonium in this kinematic region is produced  
by the hadronization of individual 
high-$p_T$ partons.   The second is the development of a factorization 
formalism for quarkonium production 
based on nonrelativistic QCD that allows the formation 
of charmonium from color-octet $c \bar c$ pairs 
to be treated systematically. 
This review summarizes these theoretical developments and their
implications for quarkonium production in high energy colliders.
\end{abstract}

\bigskip
\hrule
\bigskip

\section{INTRODUCTION}

The goal of high energy physics is to identify the elementary 
constituents of matter and to understand their fundamental interactions.
Over the last twenty years, this endeavor has been extraordinarily successful.
A gauge theory called the minimal Standard Model provides a
satisfactory description of the strong, weak, and electromagnetic interactions
of all the known elementary particles.
There are very few discrepancies between theory and experiment, and most
of them are at the level of a few standard deviations or less.
However there is 
one process for which 
experimental results have differed from theoretical 
predictions by orders of magnitude:
the production of charmonium at large
transverse momentum at the Tevatron.  
This dramatic conflict between experiment and theory
presents a unique opportunity to make a significant step forward
in  our understanding of  heavy quarkonium physics.
 
Prior to 1993, the conventional wisdom on the production  of charmonium 
in hadron collisions was based primarily on calculations in the 
{\it color-singlet model}.  In this model, the production of a 
charmonium state is assumed to proceed through parton processes 
that produce a $c \bar c$ pair in a color-singlet state.  
It is the predictions of the color-singlet model at lowest order 
in $\alpha_s$ that disagree so dramatically with the Tevatron data. 
The data can, however, be explained  
by combining 
two recent theoretical developments in heavy quarkonium physics.  
The first is the realization that heavy quarkonium 
at large transverse momentum is produced primarily 
by {\it fragmentation}, the 
hadronization of individual 
high-$p_T$ partons.
In the color-singlet model, fragmentation first contributes 
at higher order in $\alpha_s$ 
and thus was not taken into account in previous calculations.
The second development is the realization that 
{\it color-octet mechanisms}, in which the 
$c \bar c$ pair is produced at short distances in a color-octet state,
sometimes dominate the production of charmonium. 
These mechanisms can be analyzed systematically using 
a new factorization formalism for 
heavy quarkonium production that is based on an effective field theory 
called {\it nonrelativistic QCD} (NRQCD). 
The Tevatron data can be explained  by including a color-octet term
in the fragmentation function for the formation 
of charmonium in a gluon jet.  Evidence in support of this explanation 
has been accumulating, but a great deal of work, 
both experimental and theoretical, will be required
before it can be regarded as conclusive.

The purpose of this review is to  summarize the recent theoretical
developments in heavy quarkonium production.  A brief 
review of these developments 
has been given previously by Mangano \cite{mangano}.
Our main focus will be on the production of quarkonium 
in high energy colliders, where both fragmentation and color-octet 
mechanisms are important.  
The color-singlet model is reviewed in Section 2.  We summarize
its predictions for charmonium production in $p \bar p$ colliders,
which are in dramatic disagreement with Tevatron data.
The production of quarkonium by fragmentation is reviewed in 
Section 3.  We describe the factorization theorems of perturbative QCD
that imply that fragmentation should dominate at 
large transverse momentum, and we illustrate them using  
the electromagnetic production of the $J/\psi$ in $Z^0$ decay.
We then discuss the predictions of the color-singlet model for
fragmentation functions.  Color-octet mechanisms for producing 
quarkonium are reviewed in Section 4.  
We first discuss the production of P-wave states, where a 
color-octet mechanism is required for perturbative consistency.
We then describe the NRQCD factorization formalism, 
which implies that color-octet mechanisms must 
also contribute to the production of S-wave states. 
We also discuss its implications for fragmentation functions.
In Section 5, we summarize applications of the recent theoretical  
developments to the production of prompt charmonium, bottomonium,
and the $B_c$ at the Tevatron and at LEP.  
We conclude in Section 6 with an outlook on the work that will be required 
to develop a comprehensive
understanding of heavy quarkonium production in high energy processes.

\section{THE PROBLEM OF CHARMONIUM PRODUCTION}

Before 1993, most predictions for charmonium production were based
on the {\it color-singlet model}.
In this Section, we review the color-singlet model and summarize its
predictions  at leading order in $\alpha_s$
for charmonium production at large transverse 
momentum in $p \bar p$ collisions.  We then describe 
the experimental results from the Tevatron
that have forced a reexamination of the problem of charmonium production.

\subsection{\it Color-singlet Model}

It is difficult to ascribe credit for the {\it color-singlet model}, 
since many physicists were involved in its early development.
The decays of $B$ mesons into charmonium states through the process
$b \to c \bar c + s$ was first treated in the color-singlet model
by DeGrand and Toussaint, by Wise, 
and by K\"uhn, Nussinov, and R\"uckl \cite{degrand-toussaint}.
The hadronic production of $J/\psi$ through  the parton process 
$g g \to c \bar c + g$  was 
calculated by Chang \cite{chang}.  A thorough treatment of 
charmonium production in hadron collisions through $2 \to 3$ parton processes
was later presented by Baier and R\"uckl \cite{baier-ruckl}.
Guberina, K\"uhn, Peccei, and R\"uckl \cite{g-k-p-r} applied 
the color-singlet model to charmonium production 
from the decay $Z^0 \to c \bar c + \gamma$. 
Berger and Jones \cite{berger-jones} 
calculated the rate for  photoproduction of charmonium, 
and emphasized the kinematical restrictions on the applicability 
of the color-singlet model.
This model was also applied to the inclusive production of 
charmonium in $e^+e^-$ annihilation by Keung  
and by K\"uhn and Schneider \cite{keung}.
A thorough review of the applications of the color-singlet model
to heavy quarkonium production was recently given by 
Schuler \cite{schuler}.

An alternative model for quarkonium production called the 
{\it color-evaporation model} was developed around the same time 
\cite{fritzsch}.  In this model, it was assumed that all $c \bar c$ pairs 
with invariant mass between  $2 m_c$ and the $D \overline D$
threshold $2 m_D$ produce charmonium states.  The fraction $f_H$
of the $c \bar c$ pairs that form a particular charmonium state $H$
is assumed to be independent of the production process.  This model
is incapable of describing the variation of the production ratios for 
charmonium states between processes and as functions of
kinematical variables. It will therefore not be considered further.

To motivate the color-singlet model,
we can think of the production of charmonium as proceeding in two steps.
The first step is the production of a $c \bar c$ pair, 
and the second step is the binding of the $ c \bar c$ pair into a 
charmonium state.  In order to have a significant probability of binding,
the $c \bar c$ pair must be produced with relative momentum 
that, in the $c \bar c$ rest frame, is small compared to the mass 
$m_c$ of the charm quark.  Otherwise, the $c$ and $\bar c$ will
fly apart and ultimately form $D$ and $\overline D$ mesons.

We first consider the production of the $c \bar c$ pair. 
Assuming that the $c$ and $\bar c$ are not present in the initial state,
any Feynman diagram for the production of a $c \bar c$ pair must involve 
virtual particles that are off their mass shells by amounts 
of order $m_c$ or larger. 
The part of the amplitude in which all internal lines are 
off-shell by amounts of order $m_c$ or larger is called the 
{\it short-distance} part, and it is calculable 
using perturbation theory in $\alpha_s(m_c)$. 
The parts of the amplitude in which the $c$ and
$\bar c$ lines are off-shell by amounts  much less than $m_c$
can be considered part of the amplitude for the
formation of the bound state.
The short-distance part of the amplitude describes
the production of a $c \bar c$ pair
with a spatial separation that is of order $1/m_c$ or smaller.
This follows from the fact that the short-distance part 
is insensitive to changes in the relative 3-momentum of the
$c$ and $\bar{c}$ that are much less than $m_c$.
Since $1/m_c$ is much smaller than
the length scale associated with the charmonium wavefunction,
the $c \bar c$ pair is essentially pointlike on that scale.  Thus
we need only consider the amplitude for a pointlike $c \bar c$ pair to bind 
to form a charmonium state.  This amplitude will necessarily
depend on the charmonium state $H$ and
on the quantum numbers of the $c \bar c$ pair.

For any given charmonium state, the dominant Fock state consists of 
a color-singlet $c \bar c$ pair in a definite angular-momentum state.
We denote the two possible color states of a $c \bar c$ pair
by  $\underline{1}$ for color-singlet and $\underline{8}$ for color-octet.  
We use the spectroscopic notation ${}^{2S+1}L_J$
for the angular momentum state, where $S$, $L$, and $J$ are the 
quantum number associated with the total spin, the orbital angular momentum,
and the total angular momentum, respectively.
Thus the dominant Fock state for a charmonium state $H$ is denoted 
$\vert c \bar c(\underline{1},{}^{2S+1}L_J) \rangle$ for some appropriate 
values of $S$, $L$, and $J$.
For example, the dominant Fock state for the $J/\psi$ is  
$\vert c \bar c(\underline{1},{}^3S_1) \rangle$, 
while for the $\chi_{cJ}$, it is 
$\vert c \bar c(\underline{1},{}^3P_J) \rangle$. 
 
The {\it color-singlet model} is a simple model for the amplitudes
for a pointlike $c \bar c$ pair to form a charmonium state.
If the dominant Fock state of the meson $H$ is 
$\vert c \bar c(\underline{1},{}^{2S+1}L_J) \rangle$,
the amplitude is assumed to be 0 unless the pointlike $c \bar c$ pair is 
in a color-singlet ${}^{2S+1}L_J$ state.  For 
this state, the amplitude can be expressed in terms of the
$L$'th derivative of the radial wavefunction at the  origin for the meson $H$.
For example, the amplitudes for producing the states $\psi$ and $\chi_{cJ}$, 
plus some specific final state $F$, are assumed to have the forms
\begin{eqnarray}
{\cal A}(\psi + F) &=& 
\widehat{\cal A}(c \bar c(\underline{1},{}^3S_1) + F) \; R_\psi(0) ,
\label{CSM:amp-S}\\
{\cal A}(\chi_{cJ} + F) &=& 
\widehat{\cal A}(c \bar c(\underline{1},{}^3P_J) + F) \; R_{\chi_c}'(0).
\label{CSM:amp-P}
\end{eqnarray}
The $\widehat{\cal A}$'s are amplitudes for producing color-singlet 
$c \bar c$ pairs with vanishing relative momentum in the 
angular-momentum states indicated. 
The factor $R_\psi(0)$ is the radial 
wavefunction at the origin for the $\psi$, while $R_{\chi_c}'(0)$
is the derivative of the radial wavefunction at the origin for the
$\chi_c$ states.  In the color-singlet model, it is assumed 
that the amplitudes $\widehat{\cal A}$ can
be calculated using perturbative QCD and that all nonperturbative 
effects can be absorbed into the wavefunction factors.
{}From (\ref{CSM:amp-S}) and (\ref{CSM:amp-P}), we deduce that the
corresponding inclusive 
differential cross sections in the color-singlet model have the forms
\begin{eqnarray}
d \sigma (\psi + X) &=& 
d \widehat{\sigma}(c \bar c(\underline{1},{}^3S_1) + X) \;  |R_\psi(0)|^2 ,
\label{CSM:sig-S}
\\
d \sigma (\chi_{cJ} + X) &=& 
d \widehat{\sigma}(c \bar c(\underline{1},{}^3P_J) + X) \; |R_{\chi_c}'(0)|^2 .
\label{CSM:sig-P}
\end{eqnarray}

The color-singlet model has enormous predictive power.  
The cross section for
producing a quarkonium state in any high energy process is predicted in 
terms of a single nonperturbative parameter for each orbital-angular-momentum
multiplet.  For example, the nonperturbative factor is $R_\psi(0)$ 
for the S-wave states $\psi$ and $\eta_c$.  It is $R_{\chi_c}'(0)$ 
for the P-wave states $\chi_{c0}$, $\chi_{c1}$, $\chi_{c2}$, and $h_c$.
Moreover, these parameters can be determined from decays of the 
charmonium states.  For example, $R_\psi(0)$ can be determined from
the electronic width of the  $\psi$:
\begin{equation}
\Gamma(\psi \to e^+ e^-) \;\approx\; {4\alpha^2 \over 9 m_c^2} |R_\psi(0)|^2 .
\end{equation}
Thus the color-singlet model gives absolutely
normalized predictions for the production rates of charmonium states
in high energy collisions.

In spite of its great predictive power, the color-singlet model is only a
model.  There are no theorems that guarantee that the amplitude 
factors in the simple way assumed in (\ref{CSM:amp-S}) and (\ref{CSM:amp-P}).
In particular, it was never proven that higher order radiative corrections 
would respect the factored form.
In addition, the model is clearly incomplete.  For one thing, relativistic 
corrections, which take into account the relative velocity $v$ of the quark 
and antiquark, are neglected.  These corrections are probably not negligible
for charmonium, since the average value of $v^2$ is only about 1/3. 
The color-singlet model
also assumes that a $c \bar c$ pair produced in a color-octet 
state will never bind to form charmonium.  This assumption must break down
at some level, since a color-octet $c \bar c$ pair can make a 
nonperturbative transition to a color-singlet state by radiating a soft
gluon. 
The most glaring  evidence that the color-singlet model is incomplete 
comes  from the presence of infrared divergences in the 
cross sections for P-wave states.
This problem and its solution will be discussed in Section 4.  
For the moment, we simply 
note that the infrared divergence violates the factorization 
assumption implicit in (\ref{CSM:sig-P}).  It implies
that the cross section $d \widehat{\sigma}$ is sensitive to small
momentum scales, so it  cannot be calculated reliably
using perturbative QCD.

\subsection{\it Expectations for Charmonium at Large $p_T$}

Most calculations of charmonium production prior to 1993 were based on two 
crucial assumptions.  The first was that the amplitude for the formation of 
a charmonium state from a $c \bar c$ pair was accurately described by the 
color-singlet model.  The second was that the dominant production processes
for color-singlet $c \bar c$ pairs were the Feynman diagrams that
were lowest order in $\alpha_s$.  
The first thorough treatment of the 
problem of charmonium production at large transverse momentum in 
hadron-hadron collisions was given by Baier and R\"uckl in 1983 
\cite{baier-ruckl}.  In subsequent calculations \cite{halzen}, 
the contributions from $B$-meson decay into charmonium states were 
included.  The results of these calculations are summarized below.
 
We first introduce some terminology for describing charmonium production
in high energy colliders.  Charmonium is called {\it prompt}
if the point at which the charmonium state is produced 
and the collision point of the colliding beams can not be resolved using
a vertex detector.  
Prompt charmonium is produced by QCD production mechanisms. 
The decay of $b$-hadrons produces charmonium that is not prompt.
If a $b$-hadron is produced with large transverse momentum, it will
travel a significant distance before decaying weakly.
For a $b$-hadron with $p_T$ around 10 GeV,  the displacement between
the collision point and  the 
secondary vertex where the charmonium is produced  
is typically a fraction of a millimeter. 

A charmonium state that is prompt, 
but does not come from the decay of a higher charmonium state,
is called {\it direct}.
For example,  the $\psi$ is produced in the decays of 
$\chi_{c0}$, $\chi_{c1}$, $\chi_{c2}$, and $\psi'$ with branching fractions
of approximately 0.7\%, 27\%, 14\%, and 57\%, respectively.  Thus
the prompt $\psi$ signal includes direct $\psi$'s and contributions
from direct $\chi_{cJ}$ and direct $\psi'$.

A charmonium state with large momentum is called {\it isolated}
if there are no other hadrons whose momentum is nearly collinear.
Charmonium that is produced from the decay of a $b$-hadron 
with large transverse momentum is never isolated, because the remnant
hadrons from the decay of the $b$-hadron will have momentum 
that is nearly collinear.

We now consider the mechanisms for the production of charmonium 
with large $p_T$ in $p \bar p$ collisions.
First we focus on charmonium production from the weak decay of
$b$-hadrons.  Note that the decay of $b$-hadrons with large $p_T$
produces charmonium that is neither prompt nor isolated.
The dominant mechanism for producing $b$-hadrons at large
$p_T$ is the gluon fusion process $g g \to b \bar b$, followed by the 
hadronization of the $b$ or $\bar b$.  The inclusive branching fractions 
for any $b$-hadron to decay into charmonium states are presumably close to 
those for $B$ mesons.  The inclusive branching fractions 
for $B$-meson decays are approximately 1.3\% for 
$\psi$, 1\% for $\chi_{c1}$, and 0.5\% for $\psi'$.

The color-singlet model gives predictions for 
the production rate of charmonium from QCD mechanisms. 
If a  $c \bar c$ pair is produced with large $p_T$
through parton collisions, there must 
be a recoiling parton to  balance the transverse momentum.  Thus the 
production mechanisms that are of leading order in $\alpha_s$ are 
$2 \to 3$ parton processes. 
Note that the charmonium states produced by these 
processes are prompt and isolated.
In the color-singlet model, the only
$2 \to 3$ process that can produce $\psi$ or $\psi'$ is 
$g g \to c \bar c + g$.  The $2 \to 3$ processes that produce 
$\chi_{cJ}$'s with  large $p_T$ are $g g \to c \bar c + g$, 
$g q \to c \bar c + q$, $g \bar q \to c \bar c + \bar q$,
and $q \bar q \to c \bar c + g$.
The cross sections for these $2 \to 3$ 
processes are all of order $\alpha_s^3$.
At large $p_T$, the parton differential 
cross sections $d \widehat{\sigma} / d p_T^2$ scale like 
$1/p_T^6$ for the $\chi_{cJ}$ states and like $1/p_T^8$ for $\psi$ 
and $\psi'$.  Thus the parton cross sections for 
$\psi$ and $\psi'$ are suppressed relative to those for $\chi_{cJ}$ 
by a factor of $m_c^2/p_T^2$ at large $p_T$.

We now summarize the predictions of the color-singlet model
at leading order in $\alpha_s$ for  inclusive $\psi$
production in $p \bar p$ collisions \cite{halzen}.
The contribution from $b$-hadron decay falls off the most slowly with $p_T$ 
and it was predicted to dominate at the Tevatron for $p_T > $ 7 GeV. 
Of the prompt production mechanisms for $\psi$, the most important
was found to be the decay of direct $\chi_{c1}$.  
The contributions from direct $\chi_{c2}$ and direct $\psi$ 
were down by factors of about 4 and 18 at $p_T$ = 10 GeV.
Thus the conventional wisdom before 1993 was that $\psi$ production
at large $p_T$ at the Tevatron should be dominated by $b$-hadron decay,
with decays of direct $\chi_{c1}$ and direct $\chi_{c2}$ being the only other 
important mechanisms.  

In the case of $\psi'$, the decay of $b$-hadrons was predicted to be the 
only important production mechanism at large $p_T$.
It should be emphasized that these predictions were 
based on the color-singlet model and on the additional assumption that 
the dominant parton processes were of order $\alpha_s^3$.

\subsection{\it Prompt Charmonium at the Tevatron}

The first substantial data on the production of charmonium 
in $p \bar p$ collisions came from the S$p \bar p$S operating 
at a center-of-mass energy of 630 GeV.  Data collected by 
the UA1 collaboration
contained  indications of deviations from the predictions 
of the color-singlet model \cite{UA1}.  The Tevatron, which operates
at the significantly higher energy of 1.8 TeV, has provided an
opportunity to investigate charmonium production in 
much greater detail. 
The CDF collaboration
has accumulated large samples of data on 
the production of $\psi$, $\chi_{cJ}$, and $\psi'$ 
at the Tevatron.
Though analysis of the data from the 1988-1989 collider run
was hampered by the inability to separate prompt charmonium from charmonium 
that is produced by the decays of $b$-hadrons, discrepancies between
the data and the predictions of the color-singlet model were 
evident \cite{mangano:CDF}. 
However, assuming the conventional wisdom that $\psi$ production at large
$p_T$ should be dominated by the decays of 
$b$-hadrons and direct $\chi_c$'s, CDF extracted a value for the 
$b$ quark cross section from their data
on $\psi$ and $\chi_c$ production~\cite{CDF-prior}. 
This value was shown to be too large by about a factor of 2 
in the subsequent collider run~\cite{CDF-after}. 

Before the 1992-93 run of the Tevatron, CDF installed
a silicon vertex detector 
that can resolve secondary vertices separated by distances  greater
than about 10 $\mu$m from the $p \bar{p}$ collision point~\cite{SVX}. 
Prompt charmonium is produced essentially 
at the collision point, while charmonium from
the decay of $b$-hadrons is produced at a secondary vertex which is 
typically hundreds of microns away.
Thus the silicon vertex detector can be used to separate 
prompt charmonium from charmonium that is produced in $b$-hadron decays.

In the charmonium data sample
collected during the 1992-93 run, it was found that only 20\% of the
$\psi$'s  and only 23\% of the $\psi'$'s  come from
the decay of $b$-hadrons~\cite{CDF-psi}.  
Furthermore, only 32\% of the $\psi$'s come 
from $\chi_c$ decay~\cite{CDF-chi}.
Thus, the majority of the $\psi$'s and $\psi'$'s must be produced
by other mechanisms, in dramatic contradiction to the conventional wisdom.  
Furthermore,  the fraction of $\psi$ and $\psi'$
from $b$-hadron decay does not increase appreciably with increasing
$p_T$, contrary to the predictions of the color-singlet model.  

An excellent review of the results on quarkonium production from the
Tevatron, and also from other experiments, has been presented by 
Sansoni~\cite{sansoni}.
CDF has measured the cross sections for prompt $\psi$ and prompt $\psi'$
production as a function of $p_T$~\cite{CDF-psi}.  The prompt
$\psi$ signal has been resolved into those $\psi$'s that come from $\chi_c$ 
decays and those that do not~\cite{CDF-chi}.
In addition, the ratio of the production rates of $\chi_{c1}$ and 
$\chi_{c2}$ has been measured.  
The most recent CDF data on the production of prompt $\psi$'s that do not come 
from $\chi_c$ decay, prompt $\psi$'s from $\chi_c$
decay, and prompt $\psi'$'s, are shown in 
Figures 1, 2, and 3, respectively. 
The prompt $\psi$'s that do not come from $\chi_c$ decay include
direct $\psi$'s and $\psi$'s from the decay of direct $\psi'$'s. 
The predictions of the color-singlet model at 
leading order in $\alpha_s$ are shown as dashed lines. These
predictions fall several orders of magnitude below the data at large
$p_T$.  Thus the predictions of the color-singlet model at lowest 
order in $\alpha_s$ fail dramatically when confronted with the data
on charmonium production from the Tevatron.

\section{FRAGMENTATION}

The first major conceptual advance in the recent revolution in heavy
quarkonium production was the realization that {\it fragmentation} dominates
at sufficiently large transverse momentum.  Fragmentation is the formation 
of a hadron within a 
jet produced by a parton (quark,
anti-quark, or gluon) with large transverse momentum. 
As the parton emerges from the collision point, it
radiates gluons and other partons, most of which are almost collinear.
The partons  ultimately coalesce into  the hadrons that make up the jet.
In the case of production of charmonium by fragmentation,
the ``jet'' containing the charmonium state may not qualify as
a jet by conventional experimental definitions. 
For example, if most of the 
momentum of the ``jet'' is carried by the charmonium state, it may not 
satisfy a jet criterion that requires a specified number of tracks 
above a certain momentum threshold.

The word {\it fragmentation} is sometimes used to describe
the coalescence of partons into hadrons, whether or not
these partons make up a jet.  
We prefer to use the word {\it hadronization} 
to describe this general process.   The word {\it fragmentation} will be 
reserved specifically for the formation of a hadron within the jet
produced by a high-$p_T$ parton. 
Thus fragmentation involves the hadronization of the partons in the jet, 
but hadronization  also occurs in low energy processes that have nothing 
to do with jets.  
Fragmentation is a useful concept because 
the probability for the formation of a hadron within a jet is 
universal, {\it i.e.} it is independent 
of the process that produces the parton that initiates the jet.

In this Section, we introduce the factorization theorems of perturbative
QCD that guarantee that inclusive hadron production at large transverse
momentum is dominated by fragmentation.
We illustrate the factorization theorems using the simple
example of the electromagnetic production of  $\psi$ in $Z^0$ decay.
Finally we discuss the color-singlet model predictions for the 
fragmentation functions of heavy quarkonium.

\subsection{\it Factorization Theorems of Perturbative QCD}

One of the classic factorization theorems of perturbative QCD \cite{theorems} 
guarantees that inclusive hadron production 
in $e^+ e^-$ annihilation at sufficiently large energies 
is dominated by fragmentation.
After describing this factorization theorem, we discuss its 
extension to inclusive hadron production at 
large transverse momentum in hadron-hadron collisions. 

We consider the inclusive production of a hadron $H$ with energy $E$
in $e^+ e^-$ annihilation at large center-of-mass energy $\sqrt{s}$.
We are interested in the cross section in the {\it scaling limit} in 
which $E, \sqrt{s} \to \infty$ with $E/\sqrt{s}$ held fixed.
The production of the hadron also involves lower momentum scales, 
such as the hadron mass and 
the scale $\Lambda_{QCD}$ associated with nonperturbative effects in QCD.
The factorization theorem states that the cross section 
in the scaling limit has the form \cite{theorems}
\begin{eqnarray}
d \sigma (e^+e^- \to H(E) + X) && 
\nonumber \\
&& \hspace{-1.5in}
\;=\; 
\sum_i \int_0^1 dz \; d \hat{\sigma} (e^+e^- \to i(E/z) + X,\mu) \;
	D_{i \to H}(z,\mu),
\label{fact:ee-H}
\end{eqnarray}
where the sum is over parton types $i$ and the integral is over the 
longitudinal momentum fraction $z$ of the hadron $H$ relative to 
the parton $i$. 
In (\ref{fact:ee-H}), $d \hat{\sigma}$ is the differential
cross section for producing a 
parton $i$ with total 
energy $E/z$.  
This cross section is only sensitive to momenta on the order of $E$,
so it can be calculated using QCD perturbation theory.  The
effects of lower momentum scales can be systematically factored into 
the functions $D_{i \to H}(z,\mu)$, which are
called {\it fragmentation functions} or {\it parton decay functions}.
The fragmentation function $D_{i \to H}(z,\mu)$ gives the probability 
that the jet initiated by parton $i$ will
include a hadron $H$ carrying a fraction $z$ of the jet momentum.
The factorization theorem holds to all orders in perturbation theory.
For a light hadron $H$ whose mass is on the order of $\Lambda_{QCD}$,
all corrections to the factorization formula (\ref{fact:ee-H})
fall like powers of $\Lambda_{QCD}/E$.

The essential ingredient in the proof of the factorization
theorem was given by
Collins and Sterman \cite{collins-sterman}.  
They demonstrated that
a diagram that contributes to the inclusive cross section 
in the scaling limit can be separated into a
hard-scattering subdiagram that produces hard partons,
jet-like subdiagrams for each of the hard partons, and a soft part.
The soft part includes soft gluon lines that can
couple to any of the jet-like subdiagrams.
After summing over all possible connections of the soft gluons, 
one finds that the effects of the soft 
parts cancel, leaving a factored form for this contribution to the 
inclusive cross section.
In the case of inclusive hadron production, these ideas were used by  
Curci, Furmanski, and Petronzio
and by Collins and Soper \cite{curci-f-p}
to provide field theoretic definitions of the fragmentation functions. 

The factorization theorem (\ref{fact:ee-H})  requires the introduction
of an arbitrary scale $\mu$ that separates the large momentum scale 
$E$ from the lower momentum scales.
The parton cross sections and the fragmentation functions 
depend on the arbitrary scale $\mu$ in such a way that the cross 
section is independent of  $\mu$. 
The $\mu$-dependence 
of the fragmentation functions is given by an evolution equation of the form
\begin{equation}
\mu^2 {\partial \over \partial \mu^2} D_{i \to H}(z,\mu) \;=\; 
\sum_j \int^1_z {dy \over y} \; P_{i \to j}(z/y,\mu) \; D_{j \to H }(y,\mu).
\label{evol}
\end{equation}
The kernel $P_{i \to j}(x,\mu)$ describes the 
splitting of a parton $i$ into a parton $j$ with momentum fraction $x$,
and is calculable as a perturbation series in $\alpha_s(\mu)$.
At leading order in $\alpha_s$, these kernels are identical to the 
Altarelli-Parisi functions that govern the evolution of parton 
distributions.

The factorization theorem for inclusive hadron production in $e^+ e^-$
annihilation can be generalized to other high energy processes. 
The real power of these factorization theorems lies in the fact
that the fragmentation functions are  universal, {\it i.e.} they are
independent of the process that produces the fragmenting partons.
Thus, if the fragmentation functions for a hadron $H$
are determined from  $e^+ e^-$ annihilation data,
they can be used to predict the production rate of the hadron $H$ 
in jets produced by other high energy processes.

One of the important generalizations of the factorization theorem
is to inclusive hadron production at large 
transverse momentum $p_T$ in hadron-hadron collisions.
The proof of the factorization theorem for this process is more
difficult, because there are two hadrons in the initial state.  In fact,
such a proof has actually been carried out only for the 
simpler case of the Drell-Yan process for creating 
a muon pair~\cite{collins-s-s}.  However
there are no  
apparent obstacles to extending this proof to the case of inclusive hadron 
production at large $p_T$ \cite{theorems}.
The resulting factorization formula is 
\begin{eqnarray}
d \sigma (A B \to H(p_T) + X)
& = & \sum_{ijk} \int_0^1 \! dx_1 \; f_{j/A}(x_1) 
	\int_0^1 \! dx_2 \; f_{k/B}(x_2)
\nonumber \\
& & \hspace{-0.75in}
\times \int_0^1 \! dz \; d \widehat{\sigma}(j k \to i(p_T/z) + X) \; 
D_{i \to H}(z).
\label{fact:AB-H}
\end{eqnarray}
Long-distance effects can be factored into the fragmentation functions 
and into the parton distributions $f_{j/A}(x_1)$ and $f_{k/B}(x_2)$
for hadrons $A$ and $B$.
The restriction to large $p_T$ is necessary in order that the jets 
consisting of the remnants of hadrons $A$ and $B$ after the hard
scattering have large momentum relative to the jet
containing hadron $H$.
The formula (\ref{fact:AB-H}) should hold to all orders in perturbation theory,
with corrections falling like powers of $\Lambda_{QCD}/p_T$.
Implicit in (\ref{fact:AB-H}) are three arbitrary  scales:
the factorization scale $\mu_F$, which cancels between the parton 
distributions and $d \widehat{\sigma}$, the fragmentation scale
$\mu_{\rm frag}$, which cancels between $d \widehat{\sigma}$
and the fragmentation functions, and
the renormalization scale $\mu_R$ for the running coupling constant,
which appears in $d \widehat{\sigma}$. 
In low-order calculations, $\mu_{\rm frag}$ , $\mu_F$, $\mu_R$ 
should all be chosen on the order of $p_T/z$, the transverse 
momentum of the fragmenting parton.  

In 1993, Braaten and Yuan \cite{braaten-yuan:Swave}
pointed out that the factorization theorems 
for inclusive hadron production must apply to heavy quarkonium as well
as to light hadrons.  The only difference is that the leading corrections
fall as powers of $m_Q/p_T$, where $m_Q$ is the heavy quark mass, 
instead of powers of $\Lambda_{QCD}/p_T$.
Therefore the dominant production mechanism for 
charmonium at $p_T \gg m_c$ must be fragmentation.  
In retrospect, this statement 
may seem obvious, but fragmentation contributions were not included 
in any of the previous calculations of 
charmonium production in $p \bar p$ collisions summarized 
in Section~2.2. This can easily be seen from the $p_T$ dependence of the
parton cross sections.  In the factorization formula (\ref{fact:AB-H}),
the only momentum scale that the parton cross section $d \widehat{\sigma}$
can depend on is $p_T$.  Therefore, by dimensional analysis,
$d \widehat{\sigma}/dp_T^2$ scales like $1/p_T^4$ at large $p_T$.
The color-singlet model cross sections at leading order in $\alpha_s$
fall off much more rapidly with $p_T$:  $d \widehat{\sigma}/dp_T^2$ scales 
like $1/p_T^8$ for $\psi$ and $\psi'$
and like $1/p_T^6$ for $\chi_{cJ}$.

\subsection{\it Electromagnetic Production of $\psi$ in $Z^0$ Decay}

To illustrate the factorization theorem, we discuss the production of  
$\psi$ in $Z^0$ decay through electromagnetic interactions.
This example illustrates two important points.  First, the production 
process that is lowest order in the coupling constant does not necessarily 
dominate in the asymptotic region.  Second, the terms that do dominate 
in the asymptotic region can be factored into cross sections for producing 
partons and fragmentation functions.

The electromagnetic production of $\psi$ is particularly simple, because
the QCD interaction can be treated nonperturbatively 
by expressing the production amplitude in terms of a 
matrix element of the electromagnetic current between 
the QCD vacuum and a $\psi$: 
\begin{equation}
\langle \psi \vert \mbox{${2 \over 3}$} \bar c \gamma^\mu c \vert 0 \rangle 
\;\equiv\; g_\psi M_\psi^2 \epsilon^\mu ,
\label{gpsi}
\end{equation}
where $\epsilon^\mu$ is the polarization vector for the $\psi$
and $g_\psi$ is dimensionless.  The value of $g_\psi$ can be determined 
from the decay rate for $\psi \to e^+ e^-$:
\begin{equation}
\Gamma(\psi \to e^+ e^-) \;=\; {4 \pi \over 3} \alpha^2 M_\psi g^2_\psi.
\label{f:eq:coup}
\end{equation}
This gives the value $g^2_\psi=0.008$. 

The various contributions to electromagnetic $\psi$ production in 
$Z^0$ decay 
can be calculated using perturbation theory in the electromagnetic 
coupling constant $\alpha$.
Since this coupling constant is small, we might naively
expect the dominant production process to be the one that is lowest 
order in $\alpha$.  We must  remember, however, that there is also
another small dimensionless parameter in this problem, namely 
$M_\psi/M_Z$. The relative importance of a production process will be 
determined both by its order in $\alpha$ and by how it scales with
$M_\psi/M_Z$.  We will see that the leading-order process is suppressed
by $M_\psi^2/M_Z^2$, and that the dominant production process is actually 
higher order in $\alpha$.

The production process for $\psi$ that is lowest order in $\alpha$ is
$Z^0 \to \psi \gamma$, which proceeds at order $\alpha^2$ through the
process $Z^0 \to c \bar c + \gamma$. 
For the $c \bar c$ to form a $\psi$, the 
relative momentum of the $c$ and $\bar c$ in the $\psi$ rest frame
must be small compared to $m_c$.  This is in turn small compared to the 
momentum of the $\psi$, which is approximately $M_Z/2$.
If the relative momentum of the $c$ and $\bar c$ is neglected,
the amplitude is proportional to the 
matrix element $\langle \psi \vert \bar c \gamma^\mu c \vert 0 \rangle$, 
and it therefore can be expressed in terms of $g_\psi$.
The branching fraction for $Z^0 \to \psi \gamma$ is 
$5.2 \times 10^{-8}$ \cite{g-k-p-r}.
In the limit $M_\psi \ll M_Z$, the branching fraction is proportional to
$\alpha g_\psi^2 M_\psi^2/M_Z^2$, so it  vanishes in the 
scaling limit.  
The suppression factor $M_\psi^2/M_Z^2$ reflects the fact that the
$c$ or $\bar c$ must receive a momentum kick of order $M_Z$ from the photon
in order for the $c$ and $\bar c$ to be produced with small enough
relative momentum to form a bound state.

At next-to-leading order in $\alpha$, the $\psi$ can be produced 
electromagnetically via the decay
$Z^0 \to \psi + \ell^+ \ell^-$, which proceeds at order $\alpha^3$ through
the process $Z^0 \to c \bar c + \ell^+ \ell^-$.
The branching fraction for $Z^0 \to \psi + \ell^+ \ell^-$ 
is $7.5 \times 10^{-7}$~\cite{bergstrom-r,fleming:a}.  
Thus the decay rate for this 
order-$\alpha^3$ process is an order of 
magnitude larger than the order-$\alpha^2$ process $Z^0 \to \psi \gamma$.
The reason it is larger is that the factor of $M_\psi^2/M_Z^2$ in 
the decay rate for $Z^0 \to \psi \gamma$ provides larger suppression 
than the extra factor of 
$\alpha$ in the decay rate for $Z^0 \to \psi + \ell^+ \ell^-$.

We introduce a {\it scaling limit} for the production of $\psi$ 
with energy $E$ in $Z^0$ decay.  This limit is $E,M_Z \to \infty$ 
with $E/M_Z$ and $M_\psi$ held fixed.
In the scaling limit, 
the differential decay rate for the electromagnetic production 
of $\psi$ satisfies a factorization theorem that is analogous to that in 
(\ref{fact:ee-H}):
\begin{eqnarray}
d \Gamma (Z^0 \to \psi (E)+X) && 
\nonumber \\ 
&& \hspace{-1in} \;=\; 
\sum_i \int_0^1 dz \; d \widehat{\Gamma}(Z^0 \to i(E/z)+ X,\mu) \; 
	D_{i \to \psi}(z,\mu),
\label{fact:Z-psi}
\end{eqnarray}
where the sum over partons $i$ includes photons and positive and 
negative leptons.  The factor $d\widehat{\Gamma}$
is the inclusive rate for decay into 
a parton $i$ with energy $E/z$.  
The factor $D_{i\to \psi}(z)$ is a fragmentation function that
gives the probability for an electromagnetic jet initiated
by the parton $i$ to include a $\psi$ that carries a fraction $z$ of the
jet momentum.

Fleming \cite{fleming:b} pointed out that, since the differential 
decay rate for the process
$Z^0 \to \psi + \ell^+ \ell^-$ does not vanish
in the scaling limit, it must be expressible 
in the form (\ref{fact:Z-psi}). 
He found that, at leading order in $\alpha$ 
and in the scaling limit, it can be written  as  
\begin{eqnarray}
\frac{d \Gamma}{dz} (Z^0 \to \psi(z) + \ell^+ \ell^-) &=& 
2  \; \Gamma(Z^0 \to \ell^+ \ell^-) \; D_{\ell \to \psi}(z, \mu)\; 
\nonumber \\ 
&& \hspace{-0.5in}
\;+\; {d \widehat{\Gamma} \over d z}(Z^0 \to \gamma(z) 
	+ \ell^+ \ell^-, \mu) \; 
	P_{\gamma \to \psi} \;,
\label{fact0:Z-psi}
\end{eqnarray}
where $z = 2 E/M_Z$. 
The photon fragmentation probability 
is $P_{\gamma \to \psi}= 4 \pi \alpha g^2_\psi = 7 \times 10^{-4}$.
The lepton fragmentation function is
\begin{equation}
D_{\ell\to\psi}(z,\mu) \;=\;  
2 \alpha^2 g^2_\psi
\left[ {(z-1)^2+1 \over z} \log {z \mu^2 \over M_\psi^2}  -  z \right],
\label{D:ell-psi}
\end{equation}
where $\mu$ is an arbitrary factorization scale that separates 
the scales $M_Z$ and $M_\psi$.  The $\mu$-dependence of the fragmentation 
function cancels against that of $d \widehat{\Gamma}/dz$ in (\ref{fact0:Z-psi}).
Note that all the effects of QCD, both perturbative and nonperturbative, 
are absorbed into the factor $g_\psi^2$ in $P_{\gamma \to \psi}$ and
in (\ref{D:ell-psi}).   
In  the decay rate for $Z^0 \to \psi + \ell^+ \ell^-$, 
the absence of kinematic suppression factors like $M_\psi^2/M_Z^2$ 
can be attributed to
the fact that the $c \bar c$ pair that forms the $\psi$ is produced at a 
momentum scale of order $m_c$ rather than $M_Z$.
The large momentum scale $M_Z$ only enters in the decay of the $Z_0$ into 
the partons $\ell^+$, $\ell^-$, and $\gamma$.

\subsection{\it Fragmentation Functions in the Color-singlet Model}

In 1993, Braaten and Yuan pointed out that the 
color-singlet model gives predictions for the fragmentation functions for 
the formation of heavy quarkonium in  quark and gluon jets
\cite{braaten-yuan:Swave}.
They calculated  the fragmentation functions explicitly 
for $g \to \eta_c,\psi$  to leading order in $\alpha_s$.  
For example, the fragmentation function for $g \to \eta_c$ is 
calculated from the parton process $g \to c \bar c + g$ 
and is of order $\alpha_s^2$: 
\begin{eqnarray}
D_{g \to \eta_c}(z, 2 m_c) &&
\nonumber \\
&& \hspace{-1.in}
\;=\;  {\alpha_s^2(2 m_c) |R_{\eta_c}(0)|^2 \over 24 \pi m_c^3}
	\left[ 3z - 2 z^2 + 2 (1-z) \log(1-z) \right] .
\label{D:g-eta}
\end{eqnarray}
This is the fragmentation function at an initial scale of order $2 m_c$. 
It can be evolved up to a higher scale $\mu$ by using the 
evolution equations (\ref{evol}).
The fragmentation function for $g \to \psi$ 
is calculated from the parton process $g \to c \bar c + gg$ 
and is of order $\alpha_s^3$.
The fragmentation functions for 
$c \to \eta_c,\psi$ \cite{b-c-y:Z0} and $b \to B_c, B_c^*$ \cite{b-c-y:Bc}
were subsequently calculated by Braaten, Cheung, and Yuan. 
The fragmentation functions for $b \to B_c, B_c^*$ had actually been
calculated earlier by Chang and Chen \cite{chang-chen:a},
although they did not note the  universality of these functions.

The fragmentation functions for P-wave quarkonium states
have also been calculated in the color-singlet model to leading order 
in $\alpha_s$.  These fragmentation functions include $g \to \chi_{cJ}$ 
\cite{braaten-yuan:Pwave,ma:a},  
$c \to \chi_{cJ}$, and $b \to $ P-wave $B_c$ \cite{chen,yuan,ma:b}.
Some of the fragmentation functions for D-wave states have also been 
calculated \cite{cho-wise:a,cheung-yuan:dwave}.
The color-singlet model fragmentation functions for $g \to \chi_{cJ}$
are calculated from the parton process $g \to c \bar c + g$ and
are logarithmically infrared-divergent at leading order in $\alpha_s$.
The solution to this problem will be discussed in Section 4.
It should be noted that there is a discrepancy between the two
calculations of the fragmentation
functions for $g \to \chi_{cJ}$ that has not yet been resolved
\cite{braaten-yuan:Pwave,ma:a}.

The color-singlet model fragmentation functions can be resolved
into the contributions from each of the possible values of the helicity
$h$ of the charmonium state.   
The fragmentation functions for $c \to \psi_L,\psi_T$, 
where $\psi_L$ and $\psi_T$ represent the
longitudinal ($h=0$) and transverse ($|h|=1$) 
polarization components of the $\psi$, were first calculated by Chen \cite{chen}
and by Falk et al \cite{f-l-s-w}.
The fragmentation functions for $g \to \chi_{cJ}$ \cite{ma:a,c-t-w},
$b \to B_c^*$ \cite{chen,cheung-yuan:bcpol}, and $b \to$ P-wave $\bar b c$ 
states \cite{chen,yuan} have also been resolved
into the contributions from the individual helicities.
 
In the method developed by Braaten and Yuan, the fragmentation functions were 
extracted by taking the scaling limit of color-singlet model
cross sections and expressing them in the factored form demanded by
the QCD factorization theorems. This method would be rather cumbersome 
beyond leading order in $\alpha_s$.  The fragmentation functions
can also be calculated directly from the field theoretic definitions
\cite{ma:a,ma:b,ma:c}.  This method provides a significant advantage 
for calculating 
fragmentation functions beyond leading order in $\alpha_s$.
Higher order corrections may be particularly important if the leading order 
calculation gives a soft fragmentation function that is small 
for $z$ near 1~\cite{ehv}.

\section{COLOR-OCTET MECHANISMS}

The second major conceptual advance in the recent revolution in 
heavy quarkonium production is the realization that {\it color-octet
mechanisms} can be important.  Contrary to the basic 
assumption of the color-singlet model, a $c \bar c$ pair that is produced
in a color-octet state can bind to form charmonium.  While a $c \bar c$ pair
that forms charmonium must ultimately be in a color-singlet state,
the short-distance part of the process can involve the production of a
color-octet $c \bar c$ pair.

In this Section, we first discuss the case of P-wave states, where
perturbative consistency requires that a color-octet term be 
added to the color-singlet model cross section.
We then discuss a general factorization formalism for quarkonium
that is based on NRQCD. This formalism can be used to factor
production cross sections into short-distance parts
that can be computed using perturbative QCD and nonperturbative 
NRQCD matrix elements.  Color-octet  mechanisms
for S-wave states arise naturally in this framework.   Finally, we discuss
the implications of this formalism for fragmentation functions.

\subsection{\it Color-octet Mechanism for P-waves}

The most glaring  evidence that the color-singlet model is incomplete 
comes  from the presence of infrared divergences in the production
cross sections for P-wave states.  There are analogous divergences in the 
annihilation decay rates for P-wave states, which were discovered by 
Barbieri and collaborators as early as 1976 \cite{barbieri}.
They found  that the decay rate for $\chi_{cJ} \to q \bar q g$
depends logarithmically on the minimum energy of the final state gluon.
In phenomenological treatments of the annihilation decay rates of 
P-wave states, the infrared singularity was avoided by imposing an ad hoc 
infrared cutoff on the gluon energy.  This cutoff was sometimes
identified with the binding energy of the charmonium state, but 
without any good justification.

The presence of infrared divergences in the color-singlet model 
is not a problem if the divergences can be absorbed into the 
nonperturbative factor
in the cross section.  For example, there are linear infrared divergences 
associated with the Coulomb singularity in radiative corrections to 
S-wave cross sections, but they can be factored into $|R(0)|^2$.
Similarly, the radiative corrections for P-waves include a linear infrared 
divergence that can be factored into $|R'(0)|^2$.
However, the structure of the logarithmic infrared divergences for P-waves 
is such that they cannot be factored into $|R'(0)|^2$.

The solution to this problem was given by Bodwin, Braaten, and Lepage 
in 1992 \cite{b-b-l:Pwave}. They noted that 
the infrared divergence in the annihilation rate for
$\chi_{cJ} \to q \bar q g$ arises when the final state gluon becomes soft. 
After radiating the soft gluon from 
either the $c$ or the $\bar c$ in the color-singlet $^3P_J$ bound state, 
the $c \bar c$ pair is in a color-octet $^3S_1$ state. It  
then annihilates through the process $c \bar c \to q \bar q$.
In this region of phase space, the short-distance part of the decay amplitude
is therefore the annihilation of a $c \bar c$ pair in a color-octet 
$^3S_1$ state.
Thus, in addition to the conventional term in the color-singlet model,
the factorization formula for the decay rate of the $\chi_{cJ}$ 
into light hadrons must include a second term.
In the color-singlet model term, the
short-distance factor is the annihilation rate of a 
$c\bar{c}$ pair in a color-singlet $^3P_J$ state and the long-distance
nonperturbative factor is proportional to $|R'_{\chi_c}(0)|^2$.  
In the second term, the short-distance factor is the annihilation rate of a 
$c \bar c$ pair in a color-octet $^3S_1$ state and the long-distance factor is
the probability for the $\chi_{cJ}$ to contain a pointlike $c\bar{c}$
pair in a color-octet $^3S_1$ state. 
The color-octet term can be interpreted as a contribution
to the decay rate from the $c \bar c g$ component of the 
$\chi_{cJ}$ wavefunction. 

The solution to the problem of infrared divergences in the production 
cross sections for the $\chi_{cJ}$ states is similar \cite{b-b-l-y}. 
The infrared divergence arises from the radiation of a soft gluon
from the $c$ or $\bar c$ that form the color-singlet $^3P_J$ bound state.
Before the radiation, the $c \bar c$ pair is in a color-octet $^3S_1$
state.  The short-distance part of the process in this region of phase space
is the production of a $c \bar c$ pair with small relative 
momentum in a color-octet $^3S_1$ state. 
The factorization formula for P-wave cross sections must therefore 
include two  terms:
\begin{eqnarray}
d \sigma (\chi_{cJ} + X) &=&
d \widehat{\sigma}(c \bar c(\underline{1},{}^3P_J) + X) \; 
	 |R'_{\chi_c}(0)|^2
\nonumber \\
&& \;+\; (2J+1) \; 
d \widehat{\sigma}(c \bar c(\underline{8},{}^3S_1) + X) \;
	\langle {\cal O}^{\chi_c}_8 \rangle .
\label{NR:sig-P}
\end{eqnarray}
The first term is the conventional term of the color-singlet model.
In this term,
the short-distance factor is the cross section
for producing a $c \bar c$ pair with small relative 
momentum in a color-singlet $^3P_J$ state. 
The second term in (\ref{NR:sig-P}) represents a contribution to the 
cross section from a {\it color-octet mechanism}. The short-distance 
factor in this term is the cross section for producing a 
$c \bar c$ pair with small relative momentum in a color-octet $^3S_1$ state.
The long-distance nonperturbative factor $\langle {\cal O}^{\chi_c}_8 \rangle$ 
is proportional to the probability for a pointlike $c \bar c$ pair 
in a color-octet $^3S_1$ state to bind and form a $\chi_c$.
The factorization formula (\ref{NR:sig-P}) has been  
applied to $\chi_{cJ}$ production in 
$B$-meson decays \cite{b-b-l-y}, in $\Upsilon$ decays \cite{trottier},
and in photoproduction \cite{ma:photo}. 

Both the color-singlet factorization formula (\ref{CSM:sig-S}) for S-waves
and the factorization formula (\ref{NR:sig-P}) for P-waves
hold to all orders in perturbation 
theory in the nonrelativistic limit. This is the limit in which the 
typical velocity $v$ of the heavy quark in the bound state goes to 0.
Unlike the case of S-waves, a color-octet production mechanism is 
required by perturbative consistency in the case of P-wave states. 
 
This can be understood from the NRQCD factorization approach discussed
in Section 4.2.  This approach indicates that 
the color-singlet term for P-waves is suppressed 
by a factor of $v^2$ compared to S-waves. 
It is this suppression of the color-singlet term that 
makes it necessary to include the color-octet term in the case of P-waves.

\subsection{\it NRQCD Factorization}

The simple factorization formulas (\ref{CSM:sig-S}) 
for S-waves and (\ref{NR:sig-P}) for P-waves
are correct in the nonrelativistic limit, {\it i.e.} in the limit $v \to 0$, 
where $v$ is the typical relative velocity of the $c$ and $\bar c$ 
in charmonium.  Potential model calculations indicate that $v^2$ is only
about ${1 \over 3}$ for charmonium and about ${1 \over 10}$ 
for bottomonium.  Thus the neglect of relativistic corrections suppressed 
by powers of $v^2$ introduces a systematic theoretical error that may not 
be negligible.  
For specific processes, the problem may be worse than suggested by the 
magnitude  of $v^2$.  If there are other small parameters in the problem,
such as $\alpha_s(m_c)$ or kinematical parameters such as $m_c^2/p_T^2$,
the terms that are of leading order in $v^2$ may be suppressed by
these other parameters.  If this suppression is large enough,
terms that are subleading in $v^2$ may actually dominate.

The key to understanding the structure of the relativistic corrections to
quarkonium production is to unravel the momentum scales in the problem.
If there is a very large momentum  scale that is set by the kinematics 
of the production process, such as $M_Z$ in the case of $Z^0$ decay or 
$p_T$ in the case of production at large transverse momentum in 
$p \bar p$ collisions, this scale can be removed from the problem using the 
factorization theorems of perturbative QCD discussed in Section 3.
The next largest momentum scale is the heavy quark mass $m_c$,
which is certainly important in the  production of a $c \bar c$ pair.  
The actual formation of the bound state from the $c \bar c$ pair
involves smaller momentum scales, including the typical momentum $m_c v$
of the heavy quark in quarkonium, its typical kinetic energy $m_c v^2$,
and the scale $\Lambda_{QCD}$ of generic nonperturbative effects 
in QCD. In charmonium and bottomonium,
the mass $m_Q$ is large enough that effects associated with this scale
should be calculable using perturbation theory in the running coupling 
constant $\alpha_s(m_Q)$.  Nonperturbative effects become important 
at smaller momentum scales.

A powerful tool for isolating the effects of the scale $m_Q$ is 
{\it nonrelativistic QCD} (NRQCD), an effective field theory developed by
Caswell and Lepage \cite{caswell-lepage}.
NRQCD is a formulation of QCD in which the heavy quark and antiquark 
are treated nonrelativistically.  They are described by  a Schroedinger
field theory with separate 2-component spinor fields, 
rather than a single 4-component Dirac field.
The gluons and  light quarks are described by the
relativistic lagrangian for ordinary QCD.

NRQCD can accurately describe the physics of heavy quarkonium at 
energies that are much less than $m_Q$ above the rest mass.
This follows from the fact that virtual states with energies 
of order $m_Q$ or larger have lifetimes of order $1/m_Q$ or smaller.
In this time, quanta can propagate over distances that are at most of order
$1/m_Q$.  Since this distance is much smaller than the  size of the
quarkonium bound state, the effects of these virtual states
can be taken into account through local interaction terms in the 
NRQCD lagrangian.

The inverse of the mean radius of 
 
a quarkonium state is a dynamically generated momentum scale 
that is much smaller than its rest mass.  
The ratio of these momentum scales
defines a small parameter $v$ which can be identified with the typical
velocity of the heavy quark in the quarkonium state.  This small 
parameter $v$, which is defined nonperturbatively, 
can be exploited to  organize calculations of
heavy quarkonium observables into expansions in powers of $v^2$.
NRQCD is equivalent to full QCD in the sense that 
the parameters in the NRQCD lagrangian can be tuned so that its predictions
agree with those of full QCD to any desired order in $v^2$ \cite{l-m-n-m-h}.

Bodwin, Braaten, and Lepage \cite{b-b-l:NRQCD} have recently developed 
a theoretical framework for inclusive quarkonium production 
that allows relativistic 
corrections to be included to any desired order in $v^2$. 
This framework provides a factorization formula for inclusive cross sections 
in which all effects of the scale
$m_c$ are separated from the effects of lower momentum scales,
including $m_cv$. 
The derivation of this factorization formula involves two steps.
This first step is the topological factorization of diagrams that
contribute to the cross section.  This step is similar to the 
derivations of the  factorization theorems of perturbative QCD 
that were discussed in section 3.1.
For simplicity, we consider the case of 
charmonium production in $e^+ e^-$ annihilation at a center-of-mass 
energy $\sqrt{s}$ that is significantly larger than $2 m_c$.
We define a {\it scaling limit} by $\sqrt{s}, m_c \to \infty$  
with $\sqrt{s}/m_c$ fixed.  A diagram that contributes to the cross section 
in this limit can be separated into a hard-scattering subdiagram that
produces hard partons and a $c \bar c$ pair with small relative momentum, 
a jet-like subdiagram for each of the hard partons, a subdiagram that involves 
the $c$ and $\bar c$, and a soft part.  The soft part includes soft gluons
that can couple to the jet-like subdiagrams and to the $c \bar c$ subdiagram.
After summing over all possible connections of the soft gluons, 
the effects of the soft gluons
that are connected to the jet-like subdiagrams cancel,
leaving a factored form for this contribution to the cross section.
The derivation of
topological factorization requires the hard partons to all have large momentum 
relative to the $c \bar c$ pair.  This can presumably be 
generalized to processes in which there are hadrons in the initial state,
provided that the charmonium is produced with transverse
momentum large compared to $\Lambda_{QCD}$.

After the topo\-logical fac\-torization of the diagram,
the hard-scatter\-ing amp\-litude contains all effects of the scale $m_c$,
but it also depends on the relative momentum 
${\bf p}$ of the $c \bar c$ pair.
To complete the derivation of the factorization formula, 
the dependence on ${\bf p}$ must be  removed from the hard-scattering 
amplitude, so that all effects of the scale $m_c v$ reside
in the $c \bar c$ subdiagram.  This could be accomplished simply by 
Taylor-expanding the hard-scattering amplitude in powers of ${\bf p}$,
if it were not for the fact that this generates ultraviolet divergences 
in the $c \bar c$ subdiagram. 
This is where the power of NRQCD becomes evident. 
This effective field theory can be used to systematically unravel the scales 
$m_c$ and $m_c v$, so that all effects of the scale $m_c v$ are 
contained in the $c \bar c$ subdiagram.
Essentially, the $c \bar c$ subdiagram is defined to include 
all the parts of the diagram that are reproduced by NRQCD.
Those parts involving the $c \bar c$ pair that are not reproduced by NRQCD
must involve the scale $m_c$, and therefore can be included in  the 
hard-scattering subdiagram. 
The result is a factorization formula for the
inclusive cross section for producing a quarkonium state $H$ 
that  holds to all orders in $\alpha_s$. It has the form 
\begin{equation}
d \sigma ( H + X) \;=\; 
\sum_n d \widehat{\sigma}( c \bar c(n) + X) \; \langle {\cal O}_n^H \rangle,
\label{fact:NRQCD}
\end{equation}
where $d \widehat{\sigma}$ is the inclusive cross section for 
producing a $c \bar c$ pair in a color and angular-momentum state labeled
by $n$ and having vanishing relative momentum. 
The parton cross sections $d \widehat{\sigma}$ involve only 
momenta of order $m_c$ or larger, and therefore they can be calculated as 
perturbation expansions in $\alpha_s(m_c)$.
Since $d \widehat{\sigma}$ is insensitive to relative momenta that are 
much smaller than $m_c$, it describes the production of a $c \bar c$ pair 
with separation less than or of order $1/m_c$.  This separation is essentially
pointlike on the scale of 
a charmonium wavefunction, which is of order $1/(m_c v)$.
Thus the nonperturbative long-distance factor $\langle {\cal O}_n^H \rangle$
is proportional to the probability for a pointlike
$c \bar c$ pair in the state $n$ to form the  bound state $H$.
Note that the only dependence on the quarkonium state $H$ in
(\ref{fact:NRQCD}) resides in the factor $\langle {\cal O}_n^H
\rangle$.

The factorization formula  (\ref{fact:NRQCD})
holds only for  {\it inclusive} cross sections.  The matrix elements
$\langle {\cal O}_n^H \rangle$ are proportional to the  probabilities
for the formation of the charmonium state $H$, plus light hadrons whose 
total energy in the $H$ rest frame is of order $m_c v^2$.
Summation over these light hadronic states is essential for the separation 
of short-distance and long-distance effects.  
The methods required to derive the factorization formula
break down for exclusive cross sections.  The derivation also 
implies that factorization does not hold at the
amplitude level, 
contrary to the basic assumption of the color-singlet model.

The factors $\langle {\cal O}_n^H \rangle$ in  (\ref{fact:NRQCD})
can be expressed as vacuum matrix elements of 4-quark operators 
in NRQCD  \cite{b-b-l:NRQCD}.  
The operator ${\cal O}_n^H$ creates a 
pointlike $c \bar c$ pair in the state $n$, projects onto states 
that in the asymptotic future include the quarkonium state $H$, and 
finally annihilates the $c \bar c$ pair at the creation point, again in
the state $n$.
The matrix elements that play the most important roles in quarkonium 
production were denoted $\langle {\cal O}^H_1({}^{2S+1}L_J)\rangle$
and $\langle {\cal O}^H_8({}^{2S+1}L_J) \rangle$ in Ref. \cite{b-b-l:NRQCD}.
The operator ${\cal O}_n({}^{2S+1}L_J)$ creates and annihilates
a $c \bar c$ pair in 
the state $c \bar c(\underline{n},{}^{2S+1}L_J)$.  If the dominant
Fock state of the meson $H$ is $|c \bar c(\underline{1},{}^{2S+1}L_J) \rangle$,
the matrix element $\langle {\cal O}^H_1({}^{2S+1}L_J) \rangle$ can 
be related to the radial wavefunction.  For example, up to corrections 
of relative order $v^4$, we have
\begin{eqnarray}
\langle {\cal O}_1^\psi({}^3S_1)\rangle &=&
{9 \over 2 \pi} |R_\psi(0)|^2,
\label{O1:psi}
\\
\langle {\cal O}_1^{\chi_{cJ}}({}^3P_J)\rangle &=&
(2J+1) {9 \over 2 \pi} |R'_{\chi_c}(0)|^2.
\end{eqnarray}
If we keep only these terms in the factorization formula (\ref{fact:NRQCD}),
we recover the formulas (\ref{CSM:sig-S}) 
and (\ref{CSM:sig-P}) of the color-singlet model.
NRQCD has approximate heavy-quark spin symmetry, and this  implies
relations between color-octet matrix elements.  For example,
up to corrections of relative order $v^2$, we have
\begin{equation}
\langle {\cal O}_8^{\chi_{cJ}}({}^3S_1)\rangle 
\;=\; (2J+1) \langle {\cal O}_8^{\chi_{c0}}({}^3S_0) \rangle .
\end{equation}
Identifying 
$\langle {\cal O}_8^{\chi_{c0}}({}^3S_1) \rangle 
	= \langle {\cal O}_8^{\chi_c} \rangle$, one finds that
the corresponding term in the factorization formula (\ref{fact:NRQCD})
reproduces the color-octet term in the nonrelativistic 
factorization formula (\ref{NR:sig-P}) for P-wave production.

The factorization formula (\ref{fact:NRQCD}) is not particularly
useful in its most general form, 
since it involves infinitely many nonperturbative factors 
$\langle {\cal O}_n^H \rangle$. However, one can deduce from NRQCD
how the various matrix elements scale with $v$ for any particular 
quarkonium state $H$.  
For example, the scaling of $\langle {\cal O}^H_n({}^{2S+1}L_J)\rangle$
with $v$ is determined by the number of electric dipole and magnetic
dipole transitions that are required to go from the dominant Fock state 
of the meson $H$ to a state of the form
$|c \bar c(\underline{n},{}^{2S+1}L_J) + {\rm gluons} \rangle$.
The matrix element scales as $v^{3+2L}$, multiplied by $v^2$ for each 
electric dipole transition and $v^4$ for each 
magnetic dipole transition. 

The relative importance of the various terms in the factorization formula 
(\ref{fact:NRQCD}) is determined by the order in $v$ of the 
matrix element $\langle {\cal O}_n^H \rangle$ and 
by the order in $\alpha_s(m_c)$
of the parton cross sections $d \widehat{\sigma}( c \bar c(n) + X)$.
The NRQCD factorization framework acquires predictive power when the 
expansion (\ref{fact:NRQCD}) is truncated to some low order in $v^2$.  
If we truncate (\ref{fact:NRQCD}) to
lowest nontrivial order in $v$, regardless of the order in $\alpha_s$, 
we recover the factorization formulas (\ref{CSM:sig-S}) 
for S-waves and (\ref{NR:sig-P}) for P-waves.
However, if the cross sections $d \widehat{\sigma}$ for these
terms are suppressed,
other terms
in the factorization formula may be important.  This observation
is the basis for a suggestion by Braaten and Fleming \cite{braaten-fleming}
that {\it color-octet mechanisms} may also play an important role in
S-wave production.  The leading color-octet matrix elements for the $\psi$
are $\langle {\cal O}_8^\psi({}^1S_0) \rangle$,
$\langle {\cal O}_8^\psi({}^3S_1) \rangle$,
and $\langle {\cal O}_8^\psi({}^3P_J) \rangle$, all of which are suppressed by
$v^4$ relative to the leading color-singlet matrix element
$\langle {\cal O}_1^\psi({}^3S_1) \rangle$. 
 
If the parton cross sections $d \widehat{\sigma}$ multiplying 
the color-octet matrix elements have the same magnitudes
as the parton cross section multiplying 
$\langle {\cal O}_1^\psi({}^3S_1) \rangle$, then the corrections 
from the color-octet terms should be small.  However, if the parton 
cross section multiplying the color-singlet matrix element
is suppressed by powers of $\alpha_s(m_c)$ or by small 
kinematical parameters such as $m_c^2/p_T^2$, the color-octet
terms could dominate.  An example is the gluon fragmentation function
for producing a $\psi$, which is discussed in Section 4.3.

The implications of color-octet mechanisms for the production of
quarkonium in high energy colliders are discussed in Section 5.
There have also been many recent investigations of color-octet
production mechanisms in lower energy experiments,
including fixed target $\pi N$ and $p N$ collisions \cite{octet:pi}, 
$B$-meson decays \cite{octet:B}, $e^+ e^-$ annihilation \cite{octet:ee}, 
and photoproduction \cite{octet:photo,ko-lee-song}. 
These applications lie outside the scope of this review.

\subsection{\it Fragmentation Functions for Heavy Quarkonium}

The NRQCD factorization formula  (\ref{fact:NRQCD}) implies
that fragmentation functions for charmonium  have the general form
\begin{equation}
D_{i \to H}(z,\mu) \;=\; 
\sum_n  d_{i \to n}(z,\mu) \; \langle {\cal O}_n^H \rangle ,
\label{fact:D}
\end{equation}
where $d_{i \to n}(z,\mu)$ gives the probability for the  parton $i$
to form a jet that includes a $c \bar c$ pair in the state labeled by $n$, 
and $\langle {\cal O}_n^H \rangle$ is proportional to the probability 
for a pointlike $c \bar c$ pair in the state $n$ 
to bind to form a charmonium state $H$.
The coefficient $d_{i \to n}(z,2 m_c)$
at the initial scale $\mu = 2 m_c$
involves only momenta of order $m_c$,
and can therefore can be calculated as a perturbation expansion 
in the running coupling constant $\alpha_s(2m_c)$.
The fragmentation function can be evolved up to higher scales $\mu$
by using the evolution equations (\ref{evol}).
Note that the only dependence on the quarkonium state $H$ 
on the right side of (\ref{fact:D}) 
resides in the matrix elements. 

If the dominant Fock state for the meson $H$ is 
$|c \bar c(\underline{1},{}^{2S+1}L_J)\rangle$, the fragmentation function 
in the color-singlet model is obtained by keeping only the term
in (\ref{fact:D}) that involves $\langle {\cal O}^H_1({}^{2S+1}L_J)\rangle$.
In the case of gluon fragmentation into $\chi_{cJ}$, the coefficient 
of $\langle {\cal O}^{\chi_{cJ}}_1({}^3P_J) \rangle$ is logarithmically 
infrared divergent at leading order in $\alpha_s$.  The corresponding 
parton process is $g \to c \bar c(\underline{1},{}^3P_J) + g$, 
and the divergence arises when the final state gluon becomes soft.  
In this region of phase space, the short-distance part of the
fragmentation process is  
$g \to c \bar c(\underline{8},{}^3S_1)$, and the divergence from the 
radiation of the soft gluon should be absorbed into the matrix element
$\langle {\cal O}^{\chi_{cJ}}_8({}^3S_1) \rangle$.  The resulting 
expression for the fragmentation function at leading order in $\alpha_s$
is \cite{braaten-yuan:Pwave}
\begin{eqnarray}
D_{g \to \chi_{cJ}}(z,2 m_c) &=&
	{\alpha_s^2(2 m_c) \over m_c^5} d_J(z)  \;
	\langle {\cal O}^{\chi_{cJ}}_1({}^3P_J) \rangle
\nonumber \\
&& \;+\; 
{\pi \alpha_s(2 m_c) \over 24 m_c^3} \delta(1-z) \;
	\langle {\cal O}^{\chi_{cJ}}_8({}^3S_1) \rangle ,
\end{eqnarray}
where $d_J(z)$ is a dimensionless function of $z$. 
Since the matrix elements are the same order in $v$ and the color-octet term 
is lower order in $\alpha_s$, it can be expected to dominate.

The process $g \to c \bar c(\underline{8},{}^3S_1)$ gives  a term  
of order $\alpha_s$ in the gluon fragmentation function for any quarkonium
state $H$:
\begin{equation}
D_{g \to H}(z,2 m_c) \;\approx\; 
{\pi \alpha_s(2 m_c) \over 24 m_c^3} \delta(1-z) \;
	\langle {\cal O}_8^H(^3S_1) \rangle .
\label{D-glue}
\end{equation}
The delta function in (\ref{D-glue}) should be interpreted as a distribution
in $z$ that is peaked near $z = 1$ with a width of order $v^2$.
All other terms in the fragmentation function have coefficients
$d_{g \to n}(z)$ of order $\alpha_s^2$ or higher, or else have matrix elements 
that are higher order in $v^2$ than $\langle {\cal O}_8^H(^3S_1) \rangle$.
The importance of the term (\ref{D-glue}) in the fragmentation function
depends on the charmonium state $H$.
Surprisingly, it may be the most important term in the fragmentation
functions for the S-wave states $\psi$ and $\psi'$.
Although the matrix element $\langle {\cal O}_8^\psi(^3S_1) \rangle$
is suppressed by $v^4$ relative to $\langle {\cal O}_1^\psi(^3S_1) \rangle$,
the leading term in the coefficient of the color-singlet matrix element 
comes from the parton process $g \to c \bar c(\underline{1},{}^3S_1) + gg$
and is therefore of order $\alpha_s^3$.
The suppression of the color-singlet term by $\alpha_s^2$ may be more
effective than the suppression of the color-octet term by $v^4$.
The relative importance of the two terms can only be assessed after 
determining the value of the color-octet matrix element.
An estimate for this matrix element 
will be obtained from Tevatron data in Section 5.1.

\section{COLLIDER APPLICATIONS}

Our understanding of heavy-quarkonium production
in high energy colliders has been completely 
revolutionized by the theoretical developments described
in Sections 3 and 4.   It is clear that fragmentation contributions 
can dominate at large
transverse momentum, even if they are higher order in $\alpha_s$.
In addition, the formation of quarkonium from color-octet $Q \bar Q$ pairs
can in some cases dominate over the color-singlet contributions.
In this Section, we discuss several applications to collider physics 
in which these developments play an important role. These applications are the
production of prompt charmonium at the Tevatron,
the production of bottomonium and prompt charmonium at LEP, 
and the production of the $B_c$ in high energy colliders.

\subsection{\it Prompt Charmonium at Large $p_T$ in $p \bar{p}$ Collisions}

Much of the impetus for the recent theoretical developments in heavy 
quarkonium  production has come from Tevatron data on prompt
charmonium production 
at large $p_T$.  As discussed in Section 2.3, 
the experimentally measured rate 
at large $p_T$ is orders of magnitude greater than the predictions 
of the color-singlet model at leading order in $\alpha_s$. However,
the Tevatron 
results can be explained by a combination of the fragmentation
and color-octet mechanisms discussed in Sections 3 and 4.

In 1993, Braaten and Yuan~\cite{braaten-yuan:Swave} pointed out that 
fragmentation should be the most important charmonium production mechanism 
at sufficiently large $p_T$.  Doncheski,
Fleming, and Mangano~\cite{d-f-m} carried out the first explicit
calculation of the fragmentation contribution to prompt charmonium
production. They used the gluon and  charm quark
fragmentation functions from the color-singlet model to calculate the   
fragmentation contributions to direct $\psi$ and direct $\psi'$ production 
at large $p_T$ at the Tevatron. 
They found that, although fragmentation does indeed dominate 
over the leading-order color-singlet
contribution for $p_{T}$ greater than about 7~GeV,
the predictions still fell more than an order of magnitude 
below the CDF data for inclusive $\psi$ and $\psi'$ production.

In subsequent calculations of prompt $\psi$ production \cite{b-d-f-m},
contributions from fragmentation into direct $\chi_{c}$,
followed by the radiative decay of $\chi_{cJ}$ into $\psi$,  
were included.  These  contributions are roughly an 
order of magnitude larger than the direct $\psi$ contributions, 
and are dominated by the 
color-octet term in the gluon fragmentation function for $\chi_{cJ}$. 
The value of the NRQCD matrix element 
$\langle {\cal O}^{\chi_{c0}}_8({}^3S_1) \rangle$ 
that was used in this calculation
was estimated from data on $B$-meson decays \cite{b-b-l-y}.
The inclusion of the contribution from gluon fragmentation into $\chi_{cJ}$
brings the theoretical prediction for inclusive $\psi$ production
to within a factor of 3 of the  CDF data~\cite{CDF-psi}.  
Because of the many theoretical uncertainties that enter into the calculation,
this factor of 3 discrepancy was considered acceptable at the time.
However, in the case of prompt $\psi'$ production,
the theoretical prediction remained about a factor of 30 
below the data.
This dramatic discrepancy became known as the ``CDF $\psi'$ anomaly''.

The main difference between the prompt production of $\psi$ and $\psi'$ 
is that the $\psi$ signal is fed by direct $\chi_{cJ}$'s, while
the $\psi'$ signal is not fed by any known charmonium states.
The $\psi'$ anomaly could be explained if there
exist undiscovered charmonium states with sufficiently large branching 
fractions into $\psi'$.  Among the possibilities that have been considered
are D-wave states, higher P-wave states, and ``hybrid'' states whose
dominant Fock state is $c \bar c g$ \cite{c-t-w,close}.  The main 
difficulty with these proposals is explaining why these  states 
should have the large branching fractions into $\psi'$ that would be 
required to explain the data.  

An alternative solution to the $\psi'$ anomaly, based on the NRQCD 
factorization formalism, was proposed by
Braaten and Fleming~\cite{braaten-fleming}. They suggested 
that the dominant contribution to the production of $\psi'$ with large 
$p_T$ comes from the 
color-octet term (\ref{D-glue}) in the gluon fragmentation function
for $\psi'$.  This term represents the 
fragmentation of a gluon into a $c \bar c$ pair
in a color-octet ${}^3S_1$ state at short distances, followed
by the formation of the $\psi'$ from the $c \bar c$ pair through
nonperturbative  QCD interactions.  The probability for the formation
of the  $\psi'$ is proportional to
the NRQCD matrix element 
$\langle {\cal O}^{\psi'}_8({}^3S_1) \rangle$. 
While the normalization of this term in the cross section for $\psi'$
production depends on the undetermined matrix element 
$\langle {\cal O}^{\psi'}_8({}^3S_1) \rangle$, the $p_T$-dependence
of this contribution is predicted and found to be in 
good agreement with the CDF data~\cite{CDF-psi}.
By fitting to CDF data, one finds that the value  of the matrix element is 
$\langle {\cal O}^{\psi'}_8({}^3S_1) \rangle \approx 
0.0042 \; \mbox{GeV}^3$.  This value 
is consistent with the NRQCD prediction that it should be 
suppressed by $v^4$ with respect to the
corresponding color-singlet matrix element 
$\langle {\cal O}^{\psi'}_1({}^3S_1) \rangle 
\approx 0.573 \; \mbox{GeV}^3$.  Thus this solution to the $\psi'$ 
anomaly is at least plausible.

This proposal was given further support 
by subsequent CDF data in which the prompt $\psi$'s were 
separated into those that come from $\chi_c$ decay and those that 
do not~\cite{CDF-chi}.  The observed rate for prompt 
$\psi$'s not from $\chi_c$ decay 
was about a factor of 30 larger than the theoretical predictions
for direct $\psi$ from fragmentation in the color-singlet model.
Cacciari, Greco, Mangano, and Petrelli 
showed that this data could be explained by color-octet
gluon fragmentation \cite{c-g-m-p,cho-leibovich:1}.  As in the case of
$\psi'$, the $p_T$-dependence predicted by color-octet fragmentation
is in good agreement with the data.  The value of the color-octet
matrix element that is required to fit the data is 
$\langle {\cal O}^{\psi}_8({}^3S_1) \rangle \approx  0.014 \; \mbox{GeV}^3$.
This is small compared to the corresponding color-singlet matrix element
$\langle {\cal O}^{\psi}_1({}^3S_1) \rangle \approx 1.00 \; \mbox{GeV}^3$,
consistent with the suppression by $v^4$ predicted by NRQCD.

The simple picture that seems to emerge from the CDF data is that the 
dominant mechanism for  the production of $\psi$, $\psi'$,
or $\chi_{cJ}$ at large $p_T$ is the perturbative fragmentation of a 
gluon into a $c\bar{c}$ pair in a color-octet
${}^3S_1$ state, followed by the nonperturbative formation of charmonium 
from the $c \bar c$ pair. The differential cross
section therefore reduces to that for producing a gluon 
convoluted with a fragmentation function: 
\begin{eqnarray}
d \sigma (p \bar p \to H(p_T) + X)
& = & \sum_{jk} \int_0^1 \! dx_1 \; f_{j/p}(x_1) 
	\int_0^1 \! dx_2 \; f_{k/\bar p}(x_2)
\nonumber \\
& & \hspace{-1.25in}
\times \int_0^1 \! dz \;
	 d \widehat{\sigma}(j k \to g(p_T/z) + X,\mu_{\rm frag}) \; 
D_{g \to H}(z,\mu_{\rm frag}).
\label{fact:ppbar-H}
\end{eqnarray}
The dominant term in the gluon fragmentation function
is assumed to be the term proportional to 
$\langle {\cal O}^H_8({}^3S_1) \rangle$.
It is given by (\ref{D-glue}) at the scale $\mu_{\rm frag} = 2 m_c$,
but it should be evolved up to the scale  $\mu_{\rm frag} = p_T/z$
using the evolution equations (\ref{evol}).
Of course, a thorough analysis of charmonium production at large $p_T$
must include additional terms in the gluon fragmentation function 
that are of order $\alpha_s^2$ and higher.  It must also
include the contributions from the fragmentation of other partons, 
such as the charm quark and light quarks.

In Figures 1, 2, and 3, the predictions of this simple picture of 
charmonium production are compared with the
most recent CDF data on prompt $\psi$ not from $\chi_c$  
decay~\cite{CDF-chi}, prompt $\psi$ from $\chi_{c1}$ and $\chi_{c2}$  
decay~\cite{CDF-chi}, and prompt $\psi'$~\cite{CDF-psi}, respectively.
The predictions of the color-singlet model at leading order in $\alpha_s$ 
are shown as dashed lines and they  fall orders of magnitude below the data. 
The dotted lines in Figures 1 and 2 are  the fragmentation contributions 
calculated in the color-singlet model.  While they increase the
theoretical predictions by more than an order of magnitude at the 
largest values of $p_T$, they still fall about a factor of 30 
below the experimental measurements.  
The solid curves in Figures 1, 2, and 3 are the 
contributions from color-octet fragmentation, with the color-octet
matrix elements $\langle  {\cal O}^H_8(^3S_1) \rangle$ adjusted 
to fit the data.
The solid curves differ only in their normalizations.  The normalizations
in Figures 1, 2, and 3 are proportional to   
$\langle {\cal O}^\psi_8 ({}^3S_1) \rangle
	+ 0.57 \langle {\cal O}^{\psi'}_8 ({}^3S_1) \rangle$,
$0.27 \langle {\cal O}^{\chi_{c1}}_8 ({}^3S_1) \rangle
	+ 0.14 \langle {\cal O}^{\chi_{c2}}_8 ({}^3S_1) \rangle
	= 1.51 \langle {\cal O}^{\chi_{c0}}_8 ({}^3S_1) \rangle$,
and $\langle {\cal O}^{\psi'}_8 ({}^3S_1) \rangle$, respectively.
Their shapes agree reasonably well with the data.
The values of the matrix elements for $\psi$ and $\psi'$ are given 
above.  The value $\langle {\cal O}^{\chi_{c0}}_8 ({}^3S_1)
\rangle = 0.0076 \;\; \mbox{GeV}^3$ obtained
for the $\chi_c$ matrix element is significantly larger than the 
most recent value
obtained from $B$-meson decay~\cite{braaten-fleming}. 
In calculating these curves, we imposed a
pseudorapidity cut of $| \eta | < 0.6$ on the charmonium
state and we used the MRSD0 parton  distribution set. 
The scales $\mu_R$, $\mu_F$, and $\mu_{frag}$
were all chosen to be 
equal to the transverse momentum $p_T/z$ of the fragmenting gluon.

While the fragmentation contribution must dominate at sufficiently 
large $p_T$, the contributions to the cross section
that are suppressed by factors of $m^2_c/p^2_T$ 
may be important at the values of $p_T$ 
that are experimentally accessible.  These contributions have been 
included in a recent calculation by Cho and
Leibovich~\cite{cho-leibovich:2}.  
They calculated the cross sections
for all $2 \to 3$ parton processes of the form $i j \to c \bar c + k$
that produce $c \bar c$ pairs in color-octet states with angular 
momentum quantum  numbers ${}^3S_1$, ${}^1S_0$, and ${}^3P_J$. 
They used these to calculate the cross section for direct $\psi$ 
and $\psi'$ production, including all color-octet contributions that are 
suppressed only by $v^4$. 
By fitting the $p_T$ distributions for all $p_T$ greater than 5~GeV, 
they extracted values for the matrix elements
$\langle  {\cal O}^\psi_8(^1S_0) \rangle$ and 
$\langle  {\cal O}^\psi_8(^3P_J) \rangle$, 
as well as $\langle  {\cal O}^\psi_8(^3S_1) \rangle$.
The best fit requires a value for $\langle  {\cal O}^\psi_8(^3S_1) \rangle$
that is significantly smaller than that obtained above.
The terms involving 
$\langle  {\cal O}^\psi_8(^1S_0) \rangle$ and 
$\langle  {\cal O}^\psi_8(^3P_J) \rangle$ dominate in the range
$5 \; \mbox{GeV} < p_T < 10 \; \mbox{GeV}$.
At $p_T > 10 \; \mbox{GeV}$, the term involving
$\langle  {\cal O}^\psi_8(^3S_1) \rangle$ 
dominates and it asymptotically 
approaches the contribution from the leading color-octet term in the
gluon fragmentation function.  The difference
between the full calculation of the differential cross section and the 
fragmentation approximation is less than 20\% at $p_T =$  10 GeV.

The greatest weakness of the proposal to explain the CDF data 
by color-octet production mechanisms is that  
the color-octet matrix elements
are free parameters that can be adjusted to fit the data.
The predictive power in this proposal lies in the  fact that
the NRQCD matrix elements are universal,
and will also enter into other charmonium production processes.
The most convincing proof that the color-octet   
proposal is correct would be to show that the values of
the matrix elements determined from the CDF data 
are required to explain charmonium production 
in other high energy processes.
One possibility is $Z^0$ decay, which will be discussed in Section 5.2.

The color-octet fragmentation mechanism also gives other
predictions that can be tested experimentally.
The most dramatic prediction 
is that direct $\psi$'s and $\psi'$'s at large $p_T$ will 
have a large spin alignment. 
Cho and Wise \cite{cho-wise:b} pointed out 
that $\psi$'s and $\psi'$'s produced by color-octet gluon fragmentation
tend to inherit the transverse polarization of the nearly on-shell
gluon.  The process $g \to c \bar c(\underline{8},{}^3S_1)$
produces a $c \bar c$ pair whose total spin is transversely polarized.
The approximate heavy-quark spin symmetry of NRQCD implies that
the spin state of the $c \bar c$ pair 
will be affected very little by the nonperturbative 
QCD effects that are involved in binding the $c \bar c$ pair into 
a $\psi$.  At leading order in $\alpha_s$, the predicted 
polarization is 100\% transverse.
Beneke and Rothstein analyzed the radiative corrections 
and concluded that they can decrease the polarization by only about 10\%
\cite{beneke-rothstein}.
Cho and Leibovich \cite{cho-leibovich:2}
have shown that the dominant corrections to the 
spin alignment at Tevatron energies
come from terms in the cross section that involve
$\langle  {\cal O}^\psi_8(^1S_0) \rangle$ and 
$\langle  {\cal O}^\psi_8(^3P_J) \rangle$ and 
fall as $m_c^2/p_T^2$.  These corrections remain small at the largest 
values of $p_T$ measured at the Tevatron.
A large transverse spin alignment is therefore a robust signature
for the dominance by  color-octet fragmentation of
the production of direct $\psi$'s and $\psi'$'s at high $p_T$. 

There are other predictions of the color-octet  
fragmentation mechanism that can be
tested at the Tevatron. As pointed out by Barger, Fleming, and
Phillips, the production of two quarkonium states
with large $p_T$ will be dominated by the production of two
high-$p_T$ gluons which both fragment into quarkonium~\cite{bfp}. 
Taking into account the color-octet mechanism, they found that $\psi \psi$ 
events at large $p_T$ should be detectable at the Tevatron.
The color-octet fragmentation mechanism also gives predictions for
correlations between charmonium at large $p_T$ and the other jets 
produced by the $p \bar p$ collision, but these have not yet been explored.

The differential cross sections for the production of bottomonium 
states at the Tevatron have also been calculated 
by Cho and Leibovich \cite{cho-leibovich:1,cho-leibovich:2}.  
The predictions of
the color-singlet model fall more than an order of magnitude below the
CDF data for $p_T > 5 \; \mbox{GeV}$~\cite{CDF-upsilon}. 
The production rates can, however, be
explained by including color-octet terms in the cross section
for the states $\Upsilon(nS)$ and $\chi_{bJ}(n)$, $n=1,2,3$.
The values of the color-octet matrix elements that are required 
to fit the CDF data are consistent with expectations from NRQCD.
The calculations of Cho and Leibovich show 
that fragmentation
is not important in bottomonium production at the Tevatron.
The error from neglecting contributions that fall as $m_c^2/p_T^2$ 
is greater than 100\% even if $p_T$ is as large as 20 GeV.

Color-singlet and color-octet fragmentation 
processes in inelastic $\psi$ photoproduction at the HERA $ep$
collider have also been investigated~\cite{ko-lee-song,god-roy-srid}. 
The dominant
contributions for values of $p_T$ accessible to experimental
studies appear to be color-singlet charm quark
fragmentation and color-octet processes that are suppressed 
by $m^2_c/p^2_T$. In the color-octet contributions 
the $c\bar{c}$ pair is created at short distances in either 
a $^1S_0$ or $^3P_J$ state. Because the numerical values of the 
color-octet matrix elements $\langle {\cal O}^{\psi}_8 (^1S_0) \rangle$ 
and $\langle {\cal O}^{\psi}_8 (^3P_J) \rangle$ are not measured
very precisely it is not possible to determine
which process is the dominant one at HERA. However at asymptotic values 
of $p_T$ the fragmentation contribution will be the most important
contribution. Thus for large $p_T$ data from HERA 
can be used to study a fragmentation mechanism that is not so easily
probed at the Tevatron and at LHC.

\subsection{\it Charmonium and Bottomonium in $Z^0$ Decay}

The decay of the $Z^0$ provides a laboratory for the study 
of heavy quarkonium with large transverse momentum 
that is complementary to $p \bar p$ collisions.
The color-singlet model predicts that the production rates for
prompt charmonium and for bottomonium are so small that they are
unlikely to be observed at LEP.  However, the dramatic failure 
of the color-singlet model in $p \bar p$ collisions suggests 
that its predictions for $Z^0$ decay should also be reexamined. 

The predictions of the color-singlet model for charmonium production 
in $Z^0$ decay have been studied thoroughly 
\cite{g-k-p-r,keung,b-c-k,h-m-s}.
The largest production process for $\psi$ in the color-singlet model is
$Z^0 \to \psi c \bar c$.  The rate for this process was first
calculated in 1990 by Barger, Cheung, and Keung \cite{b-c-k}, who 
found it to be surprisingly large. It is  two orders 
of magnitude larger than the rate for $Z^0 \to \psi gg$ \cite{keung},  
despite the fact that both processes are of the 
same order in $\alpha_s$.  
This was explained in 1993 by Braaten, Cheung, and Yuan, who 
 pointed out  that the reason
$Z^0 \to \psi c \bar c$ is so large is that this process has a fragmentation 
contribution  that is not suppressed by $m_c^2 /M_Z^2$ 
\cite{b-c-y:Z0}.  They showed that the dominant terms 
can be factored 
into the decay rate for $Z^0 \to c \bar c$ and fragmentation functions
for the processes $c \to \psi c$ and $\bar{c} \to \psi \bar{c}$:
\begin{equation}
\frac{d \Gamma}{dz_\psi} (Z^0 \to \psi(z_\psi) + X) \;\approx\;
2 \; \Gamma (Z^0 \to c \bar c) \; D_{c \to \psi}(z_\psi) ,
\end{equation}
where $z_\psi = 2 E_\psi/M_Z$. 
Since the momentum scale for the fragmentation process is of order $m_c$,
the rate is proportional to $\alpha_s^2(m_c)$ instead of $\alpha_s^2(M_Z)$.
The resulting prediction for the branching fraction for direct
$\psi$ production in $Z^0$ decay is $2.9 \times 10^{-5}$. 
The experimental study of prompt charmonium in $Z^0$ decay 
is complicated by the large background  from the 
decay of $b$-hadrons, which must be removed using vertex detectors.  
There are preliminary results from LEP \cite{LEP-psi} 
that indicate that the branching 
fraction for prompt $\psi$ production is around $10^{-4}$, well
above the predictions of the color-singlet model. 

Another fragmentation process in the color-singlet model was considered 
by Hagiwara, Martin, and Stirling in 1991 \cite{h-m-s}. 
They calculated the rates for the process $Z^0 \to q \bar q g^*$,
with charmonium being produced by the virtual gluon through
the processes $g^* \to \psi g g$ and $g^* \to \chi_{cJ} g$. 
The rate is at least an order of magnitude smaller than the charm 
fragmentation process described above, but the experimental backgrounds
are less severe.  
 The DELPHI collaboration~\cite{LEP-psi} has recently 
reported $4.1 \times 10^{-4}$ as a limit on the branching fraction for 
$Z^0 \to \psi X$ from this process. 

The color-octet mechanism that was introduced to explain prompt 
charmonium production at the Tevatron also offers new possibilities
for charmonium production at the $Z^0$ resonance. 
Cheung, Keung, and Yuan \cite{c-k-y} and Cho \cite{cho} 
have studied  prompt charmonium production at LEP via 
the color-octet mechanism.
The color-octet process that is leading order in $\alpha_s$ 
involves the short-distance decay 
$Z^0 \to c \bar c(\underline{8},{}^3S_1) + g$, but it has a negligible 
branching fraction because it is suppressed by a
short-distance factor of $m_c^2/M_Z^2$.
The dominant color-octet process is of order $\alpha_s^2$ 
and involves the short-distance decay 
$Z^0 \to c \bar c(\underline{8},{}^3S_1) + q \bar q$, 
where $q$ can be one of the light quarks 
$u,d,$ or $s$ or one of the heavy quarks $c$ or $b$. 
Apart from different coupling constants, 
color factors, and long-distance matrix elements, 
this process is identical to the 
electromagnetic production of $\psi$ studied by Fleming 
\cite{fleming:b} and discussed in Section 3.2. 
At leading order in $\alpha_s$, the perturbative QCD factorization 
formula for this process has the form
\begin{eqnarray}
\frac{d \Gamma}{dz_\psi} (Z^0 \to \psi(z_\psi) + X) & \approx &
2 \sum_q \Gamma (Z^0 \to q \bar q) D_{q \to \psi}(z_\psi,\mu) 
\nonumber \\
&& \hspace{-1.in} \;+ \;
\sum_q \int_{z_\psi}^1 dy \frac{d \widehat{\Gamma}}{dz_g} 
	(Z^0 \to g(z_\psi/y) q \bar q,\mu) 
	D_{g \to \psi}(y),
\end{eqnarray}
where  $z_\psi = 2 E_\psi/M_Z$ and $z_g=z_\psi/y$.  
The fragmentation function $D_{g \to \psi}(y)$ for the formation of a $\psi$
in a gluon jet through the color-octet mechanism is given in (\ref{D-glue}),
except that it must be evolved up to the scale $E_\psi/y$ 
set by the gluon energy.
The fragmentation function  $D_{q \to \psi}(z)$  for the formation 
of a $\psi$ in a light quark jet 
can be obtained from (\ref{D:ell-psi}) by replacing $\alpha^2 g_\psi^2$ 
by $\alpha_s^2 \langle {\cal O}^\psi_1({}^3S_1) \rangle / (72 m_c^3)$.

In Figure 4, we compare the 
$\psi$ energy distribution
predicted by the  color-singlet model with the distribution from 
the most important color-octet process
$Z^0 \to c \bar c(\underline{8},{}^3S_1) + q \bar q$.   
The most striking feature in Figure 4 is that the energy 
distribution of the color-octet process dominates 
over that of the color-singlet model for all values 
of $z_\psi = 2 E_\psi / M_Z$.  
There is also a dramatic  difference between the 
shapes of the two distributions.
The energy distribution from the color-singlet process is
rather hard because of the nature of heavy quark fragmentation.
The distribution for the color-octet process is very soft and has a 
pronounced peak near  $z_\psi = 0.1$.  
The prediction for the branching ratio for $Z^0 \to \psi + X$
from the color-octet process is $1.4 \times 10^{-4}$.
This is consistent with the DELPHI bound for the 
color-singlet process $Z^0 \to \psi gg q \bar q$,
which produces $\psi$'s with a similar event topology.

The color-octet fragmentation process
is also important in the production of $\psi'$ and $\chi_c$. 
The energy distributions are predicted to be similar to that for the $\psi$ 
shown in figure 4.  The production rates are predicted to be smaller
by about a factor of 3 for the $\psi'$ and larger by about a factor of 5 for
the three spin states of the $\chi_{cJ}$ combined. 

Just as in the case of $p \bar p$ annihilation, the color-octet fragmentation 
mechanism makes definite 
predictions for the spin alignment of the $\psi$
in $Z^0$ decay. The color-octet process 
$Z^0 \to c \bar c(\underline{8},{}^3S_1) + q \bar q$
involves contributions from both gluon fragmentation 
and light quark fragmentation.  At leading order in $\alpha_s$,
gluon fragmentation only produces $\psi$'s that are transversely polarized.
However, a light quark can fragment into a longitudinally polarized $\psi$
even at leading order in $\alpha_s$. 
This leads to significant degradation of the spin alignment of 
the $\psi$.  The spin alignment varies significantly with $z$, but  
the average transverse polarization may be as small 
as 75\% \cite{c-y:spin}.

Bottomonium states can also be produced in $Z^0$ decay.  
These states  are simpler to study experimentally,
since there is no background to bottomonium production 
from non-prompt production mechanisms analogous to $b$-hadrons  
decay into charmonium. A signal from the $J^{PC} = 1^{--}$ states
$\Upsilon(1S)$, $\Upsilon(2S)$, and $\Upsilon(3S)$ has in fact been 
observed at LEP.  The preliminary result from the OPAL detector 
is that the branching fraction summed over the three states is
\cite{LEP-upsilon}
\begin{equation}
\sum_{n=1}^{3} {\rm Br}(Z^0 \to \Upsilon(nS) + X) = 
(1.2 ^{+0.9}_{-0.6} \pm 0.2) \times 10^{-4} \;\; .
\label{OPAL}
\end{equation}
The direct production mechanisms for $\Upsilon(nS)$ in $Z^0$ decay are 
essentially identical to those for prompt $\psi$.  The dominant 
color-singlet process is the decay 
$Z^0 \to \Upsilon(nS)  + b \bar b$.
This process has a fragmentation contribution
which corresponds  
to the decay $Z^0 \to b \bar b$ followed by the formation of $\Upsilon(nS)$ 
through the fragmentation of the $b$ or $\bar b$.  
The prediction for the sum of branching fraction given in (\ref{OPAL}) is 
$1.6 \times 10^{-5}$, which is about an order of magnitude below the 
OPAL result.  Color-octet mechanisms for the production of $\Upsilon(nS)$
and $\chi_{bJ}(n)$ 
have been studied by Cho \cite{cho}.  The dominant mechanism
is the formation of bottomonium through the short-distance process
$Z^0 \to b \bar b (\underline{8},{}^3S_1) + q \bar q$, $q = u,d,s$,
which  involves both  gluon fragmentation and light quark fragmentation.
Using values for  the color-octet matrix elements 
$\langle {\cal O}_8^H({}^3S_1) \rangle$ for the bottomonium states 
that were extracted from
CDF data \cite{CDF-upsilon}, Cho predicted that the branching 
fraction in (\ref{OPAL}) should be $4.1 \times 10^{-5}$.
Given the large uncertainties in the color-octet matrix elements, this is 
in reasonable agreement with the OPAL result (\ref{OPAL}).  If 
one can show that 
the same values of the color-octet matrix elements
explain $\Upsilon$ production in $p \bar p$ colliders and in $Z^0$
decay, this would provide strong support for the proposal that 
the production is dominated by color-octet mechanisms.

\subsection{\it Production of the $B_c$}

In the Standard Model, the only bound states consisting 
of two heavy quarks with different flavors 
are the $\bar b c$ mesons. 
The ground state of the $\bar b c$ system is the pseudoscalar state $B_c$.
The mass spectrum of the $\bar b c$ mesons
can be predicted reliably from quark potential models 
that are tuned to reproduce the spectra of
charmonium and bottomonium \cite{eichten-quigg}.
According to the potential-model calculations, 
the first two sets of S-wave states, the first 
and probably the entire second set of P-wave states, and 
the first set of D-wave states all lie below the $B D$ flavor 
threshold.  Since QCD interactions are diagonal in flavors, 
the annihilation of $\bar b c$ mesons can only occur 
through a virtual $W^+$  and  is therefore suppressed 
relative to the electromagnetic and hadronic transitions to  
lower-lying $\bar b c$ states.
Thus all the excited states below the $B D$ threshold  will cascade down to 
the ground state $B_c$ via emission of photons and/or pions. 
A calculation of the inclusive production of the 
$B_c$ meson must therefore include the contributions from production 
of all the S-wave, P-wave, and D-wave states that are below the 
$B D$ threshold.   

The production of the $B_c$ and the lowest $^3S_1$ state $B_c^*$ 
in $e^+e^-$ annihilation was first computed to leading order in $\alpha_s$
in the color-singlet model by Clavelli  
and by Amiri and Ji \cite{clavelli}.  
The lowest-order process for $B_c$ production is $e^+ e^- \to B_c + b \bar c$.
Chang and Chen \cite{chang-chen:a} 
showed that the energy distributions for direct $B_c$ 
and $B_c^*$ production 
could be expressed in terms of the fragmentation functions
$D_{\bar b \to B_c}(z)$ and $D_{\bar b \to B_c^*}(z)$.  
For example, the production rate for $B_c$ at the 
$Z^0$ resonance in the scaling limit
$M_Z \to \infty$ with $z = 2 E_{B_c}/M_Z$ fixed has the form
\begin{equation}
\frac{d \Gamma}{dz} (Z^0 \to B_c(z) + X) \;\approx\;
\Gamma (Z^0 \to b \bar b) \; D_{\bar b \to B_c}(z) .
\end{equation}
The fragmentation functions 
$D_{\bar b \to B_c}(z)$ and $D_{\bar b \to B_c^*}(z)$ 
were also calculated later by Braaten, Cheung, and Yuan \cite{b-c-y:Bc},
who pointed out that they were universal and could also be used to 
calculate direct $B_c$ production in other high-energy processes,
such as $p \bar p$ collisions.  The first calculations of $B_c$
production in hadron colliders using fragmentation functions
were carried out by Cheung \cite{cheung}.
The fragmentation functions were subsequently 
calculated for the P-wave states of the $\bar b c$ system
\cite{chen,yuan} and even for the D-waves \cite{cheung-yuan:dwave},
allowing a complete calculation of the inclusive $B_c$ production rate. 
The advantage of the fragmentation approach over carrying out a full
calculation 
 
of the production rate to leading order in $\alpha_s$
is that the calculation of the fragmentation function is much simpler
and it can be extended to P-wave and D-wave states relatively 
easily.   The limitation of the fragmentation approach is that it 
is accurate only  at sufficiently  large $p_T$.

Color-octet production mechanisms are expected to be much less important
for $\bar b c$ mesons than for  charmonium or bottomonium.
All terms in the gluon fragmentation functions are of order
$\alpha_s^3$ or higher. For the $b$-quark fragmentation functions,
the leading color-singlet terms and the leading color-octet terms are both of 
order $\alpha_s^2$.  For S-wave states like $B_c$ and $B_c^*$, 
the color-octet terms are therefore suppressed relative to the 
color-singlet terms by a factor of $v^4$ from the ratio of the 
NRQCD matrix elements. For P-wave states, a color-octet
term in the fragmentation function is necessary to avoid an infrared divergence
in the color-singlet fragmentation function.
However, this divergence occurs first at next-to-leading 
order in $\alpha_s$.  Thus the color-octet term is not as important as 
it is for the case of gluon fragmentation into $\chi_c$, where the divergence 
occurs at leading order.

The production of the $B_c$ and $B_c^*$ at hadronic colliders 
like the Tevatron and the  Large Hadron Collider (LHC) 
has recently been calculated by several groups, 
both through complete ${\cal O}(\alpha_s^4)$  
calculations \cite{chang-chen:b,klr:a} 
and using the simpler fragmentation 
approximation~\cite{cheung,cheung-yuan:bctev}.  
Discrepancies among the earlier calculations raised questions about 
the accuracy of the fragmentation approximation in the $\bar b c$ case. 
Recent calculations \cite{klr:a}
have shown that the fragmentation approximation 
is very accurate for the $B_c$ for $p_T$ as low as 10 GeV.
However, for the $B_c^*$, the accuracy is much lower, decreasing to 
20\% only for $p_T$ around 30 GeV.
For $B_c$ production in $\gamma\gamma$ collisions, even larger
values of $p_T$ are required before the fragmentation 
approximation becomes accurate  \cite{klr:b}. 
The fact that such large values of $p_T$ are required
in order for fragmentation to dominate can be attributed
to the presence of an additional small parameter $m_c/m_b$ in the problem.
Contributions that fall off asymptotically as
$m_b^2/p_T^2$ may dominate at sub-asymptotic values of $p_T$ 
if they are enhanced by powers of $m_b/m_c$.

The only calculations presently available for the 
production of the P-wave  $\bar b c$ mesons
have been carried out using the fragmentation approximation.
This approach has been used by Cheung and Yuan to calculate
the inclusive production of the $B_c$ in $p \bar p$ collisions,
including the contributions from all the  S-wave and 
P-wave states  below the $B D$ threshold \cite{cheung-yuan:bctev}.  
With acceptance cuts of $p_T > 6$ GeV and 
$\vert y \vert < 1$, the inclusive production 
cross section for the $B_c$ meson at the Tevatron   is about 
5 nb.  The contributions from the excited S-wave states and from the 
P-wave states  are 58\% and 23\%, 
respectively, and hence significant. 
The  D-wave states were not included in this analysis since they are 
expected to contribute only about 2\%  to 
the inclusive production of $B_c$ \cite{cheung-yuan:dwave}. 
The search for the $B_c$ is now underway at the Tevatron, and
preliminary results have 
already been presented by the CDF collaboration \cite{bcsearch-tevatron}.
If the $B_c$ is not found in the present run of the Tevatron, 
it should certainly be discovered after the
installation of the Main Injector, which will 
boost the luminosity by about a factor of 10.

The $B_c$ may also be discovered at LEP.
Candidate events for the  $B_c$
have already been reported by the ALEPH collaboration \cite{bc-lep}.
The branching fraction predicted from the color-singlet 
processes $Z^0 \to B_c + b \bar c$ and $Z^0 \to B_c^* + b \bar c$ 
 are $5.9 \times 10^{-5}$ and $8.3 \times 10^{-5}$, 
respectively \cite{b-c-y:Bc}. 
If the fragmentation functions are used to estimate the contributions
of higher S-wave, P-wave, and D-wave states, the inclusive branching 
fraction for $B_c$ is predicted to be larger by about a factor of 5.

\section{OUTLOOK}

The NRQCD factorization formalism provides a very general framework 
for analyzing the production of heavy quarkonium.
It implies that the production  process is much more complex than 
has been assumed 
in the color-singlet model.  There are infinitely many nonperturbative
matrix elements that contribute to the cross sections, 
although only a finite number of them contribute 
at any given order in $v^2$.  The most dramatic consequence of this 
formalism is that color-octet mechanisms 
sometimes give the largest contributions to the cross section. 
In the case of P-waves, there is a color-octet term  at leading order 
in $v^2$ that must be included for perturbative consistency.
In the case of S-waves,
color-octet terms are suppressed by $v^4$, but they may be important if 
the color-singlet term is suppressed by other small parameters 
such as $\alpha_s$.

The NRQCD factorization formalism suggests an explanation for the 
large production rates for bottomonium and prompt charmonium that 
have been observed at the Tevatron.  They can be attributed to 
color-octet terms in the cross section that are important because 
the color-singlet terms are suppressed by
powers of $\alpha_s$ and $m_Q^2/p_T^2$.
The magnitudes of the color-octet matrix elements that are
required to explain the Tevatron data are in accord with 
expectations from NRQCD.  This explanation leads to many predictions 
that can be tested experimentally.
The ultimate test is that the same matrix elements must be able to 
explain heavy quarkonium production in other high energy processes,
such as $Z^0$ decay. 

In the NRQCD factorization
formalism, there are many nonperturbative matrix elements 
that must be determined phenomenologically in order to make theoretical 
predictions for quarkonium production.  Fortunately, there is a wealth of 
experimental data that can be used to determine these matrix elements.
Quarkonium is produced as a byproduct of almost every high energy experiment.
In this review, we have only discussed data from the 
highest-energy colliders, the  Tevatron and LEP.  However there is also 
data on quarkonium production from lower-energy $e^+ e^-$ and $ep$
colliders.  There is also an abundance of data on $pN$, $\pi N$, 
and $\gamma N$ collisions from fixed target experiments. By carrying out a 
comprehensive analysis of all the data on quarkonium production,
it should be possible to determine most of the NRQCD matrix elements
that are of phenomenological importance.  

An enormous amount of theoretical work will be required in 
order to carry this program to completion.
For every process, the production rate should be calculated to 
next-to-leading order in all the  small 
parameters in the problem, including
$v^2$, $\alpha_s$, and kinematical parameters such as $m_Q^2/p_T^2$.
Most calculations until very recently were carried out within the 
color-singlet model, and therefore include only contributions that are 
of leading order in $v^2$.  Even in the color-singlet model, the only
process for which the corrections at next-to-leading order in 
$\alpha_s$ have been calculated is photoproduction of $J/\psi$
\cite{kramer}.  The effort to calculate contributions that are beyond 
leading order in $v^2$, such as color-octet production mechanisms,
has just begun. 

While the NRQCD factorization approach provides a very general
framework for analyzing quarkonium production, it should not be regarded 
as a complete theory.  
The derivation of the factorization formula breaks down when
there are hadrons in the initial state if the quarkonium is produced 
with small transverse momentum.
Therefore, processes involving 
diffractive scattering, such as the elastic scattering process 
$\gamma p \to \psi p$, are not described by this formalism.
The NRQCD formalism also is not appropriate for describing 
the formation of charmonium from  intrinsic $c \bar c$ pairs 
that come from the parton distribution of a colliding 
hadron \cite{vogt-brodsky}.  

The complications of diffractive scattering and intrinsic $c \bar c$
pairs do not arise in the production of heavy quarkonium at large
$p_T$, and therefore the theoretical analysis of this process 
is particularly clean.  The factorization theorems of perturbative 
QCD guarantee that the production is dominated by fragmentation,
the formation of heavy quarkonium within ``jets'' that are initiated 
by single high-energy partons.  
The large cross sections for prompt
charmonium production that have been observed at the Tevatron can
be explained by including a color-octet term in the gluon 
fragmentation function.  While the normalization of the cross section
is not predicted,
this mechanism does give other predictions that can be tested 
experimentally.  The most dramatic prediction is a 
large transverse spin alignment for direct $\psi$ and $\psi'$ at large $p_T$, 
but the spin alignment has not yet been measured.
This mechanism also gives predictions for the correlations between
high-$p_T$ charmonium states and other jets produced by the collision
that can be tested.

An enormous amount of theoretical work remains to be done in 
order to obtain precise predictions for fragmentation 
that can be compared with experiment.  Quantitative estimates of
all the relevant 
NRQCD matrix elements are required in order to determine which terms 
in the fragmentation functions are numerically important.
The important terms should all be calculated to next-to-leading 
order in $\alpha_s$.  It is also important to calculate the
power corrections to the cross sections that fall as 
$m_Q^2/p_T^2$, since they may give the largest corrections
to some observables.

New experimental data from  high-energy colliders
is providing stringent tests of our understanding of  the production of 
heavy quarkonium.   The NRQCD factorization approach suggests that 
cross sections should be calculable 
in terms of the heavy-quark mass $m_Q$, the running coupling constant
$\alpha_s$, and a few  matrix elements that can be determined 
phenomenologically.  This approach provides an explanation for 
the large production rates that have been observed at the Tevatron 
and at LEP, but further effort, 
both theoretical and experimental, will be required to show conclusively
that this explanation is correct.
If this effort is successful, it will mark a major 
milestone on the road to a comprehensive description of heavy-quarkonium 
production in all high energy processes.

We would like to thank G.T. Bodwin, Y.-Q. Chen, K. Cheung, P. Cho, 
W.-Y. Keung, G.P. Lepage, I. Maksymyk, and M.L. Mangano 
for valuable discussions. 
This work was supported in part by the U.S. Department of Energy,
Division of High Energy Physics, under the Grants  DE-FG02-91ER40690,
DE-FG02-95ER40896, and DE-FG03-91ER40674. The work of S.F. was also
supported by the University of Wisconsin Research Committee with
funds granted by the Wisconsin Alumni Research Foundation.

\vfill
\eject

\section*{Figure Captions}

\bigskip

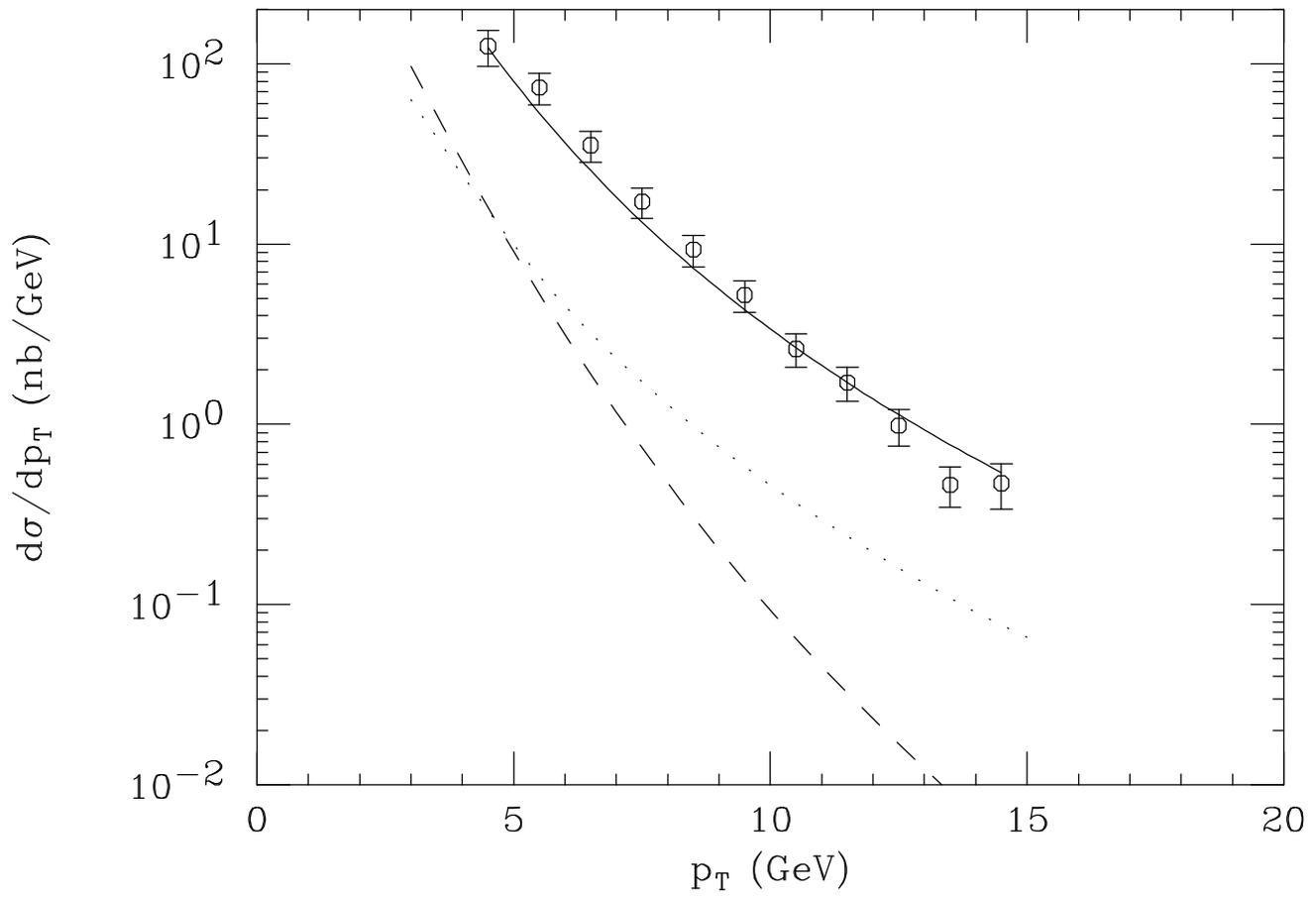
{\it Figure 1} $\;$
CDF data on the differential cross section 
for prompt $\psi$'s that do not come from 
$\chi_c$ decay as a function of $p_T$. 
The curves are the leading-order 
predictions of the color-singlet model (dashed curve), the predictions 
from fragmentation in the color-singlet model (dotted curve),
and the contribution from gluon fragmentation via the color-octet mechanism  
(solid curve) with the normalization adjusted to fit the CDF data. 
%
%

\bigskip

\noindent
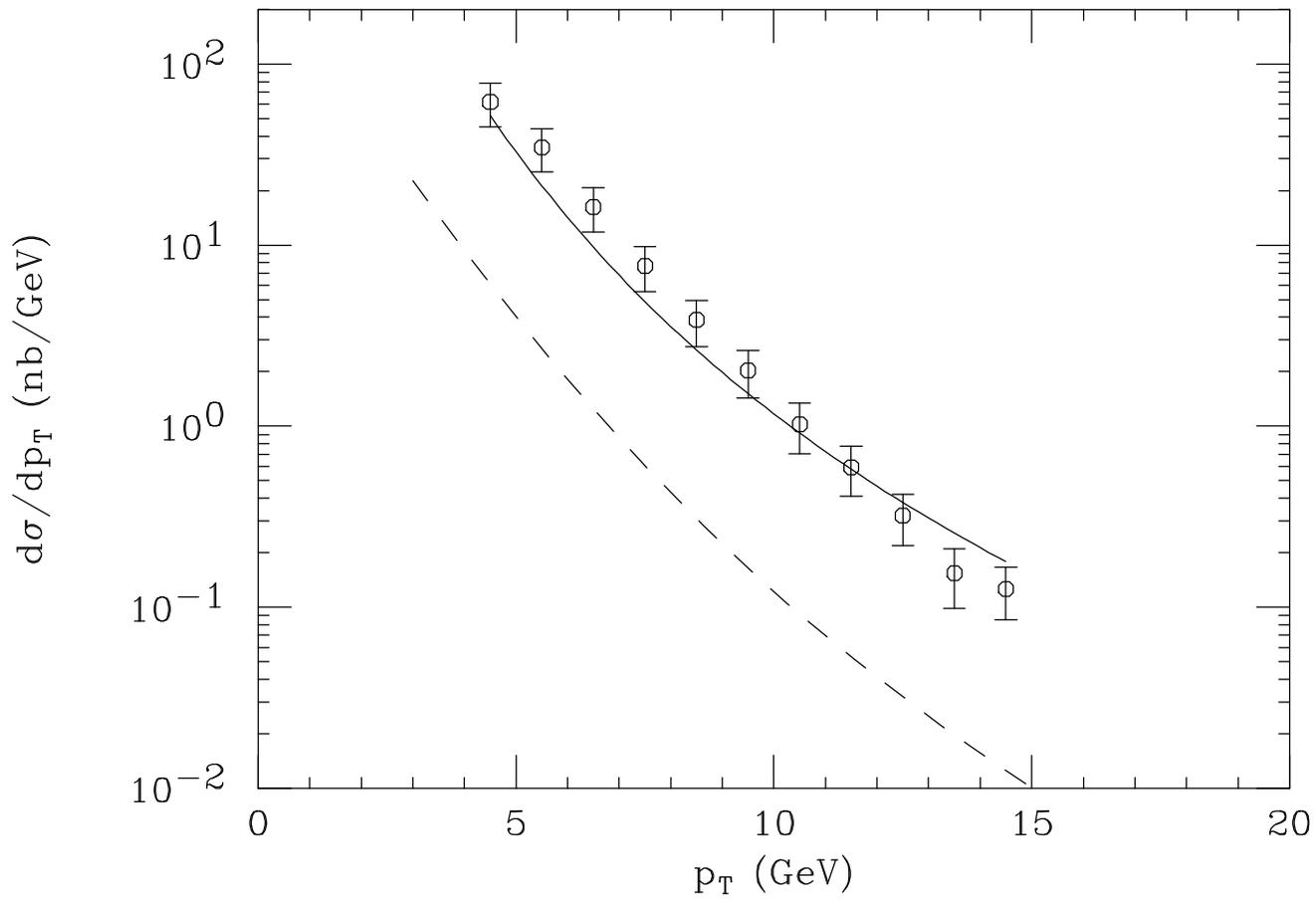
{\it Figure 2} $\;$
CDF data on the differential cross section 
for prompt $\psi$'s from $\chi_c$ decay
as a function of $p_T$.
The dashed and solid curves are as described in Figure 1.
%
%

\bigskip

\noindent
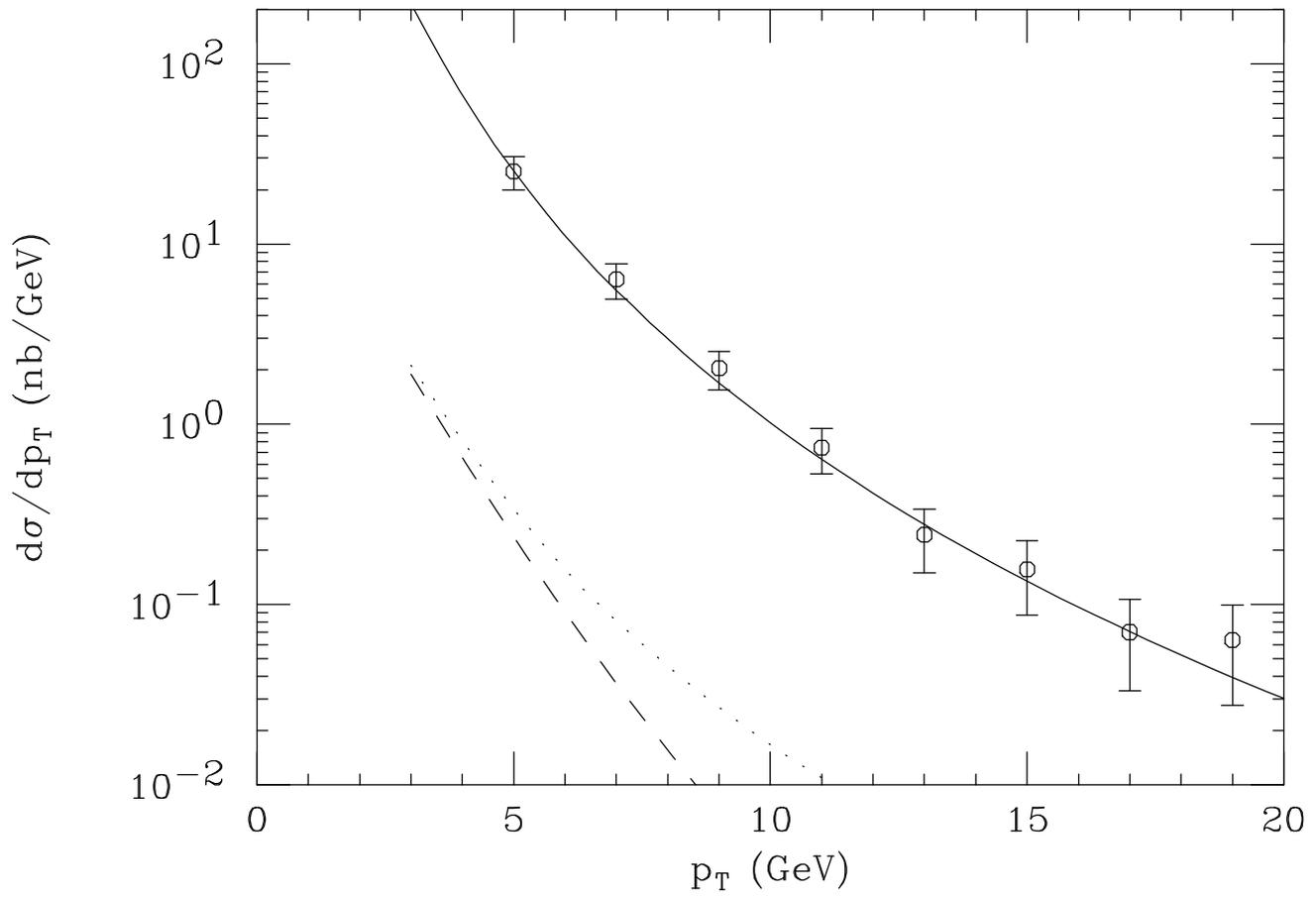
{\it Figure 3} $\;$
CDF data on the differential cross section for prompt $\psi'$'s 
as a function of $p_T$. The curves are as described in Figure 1.
%
%

\bigskip

\noindent
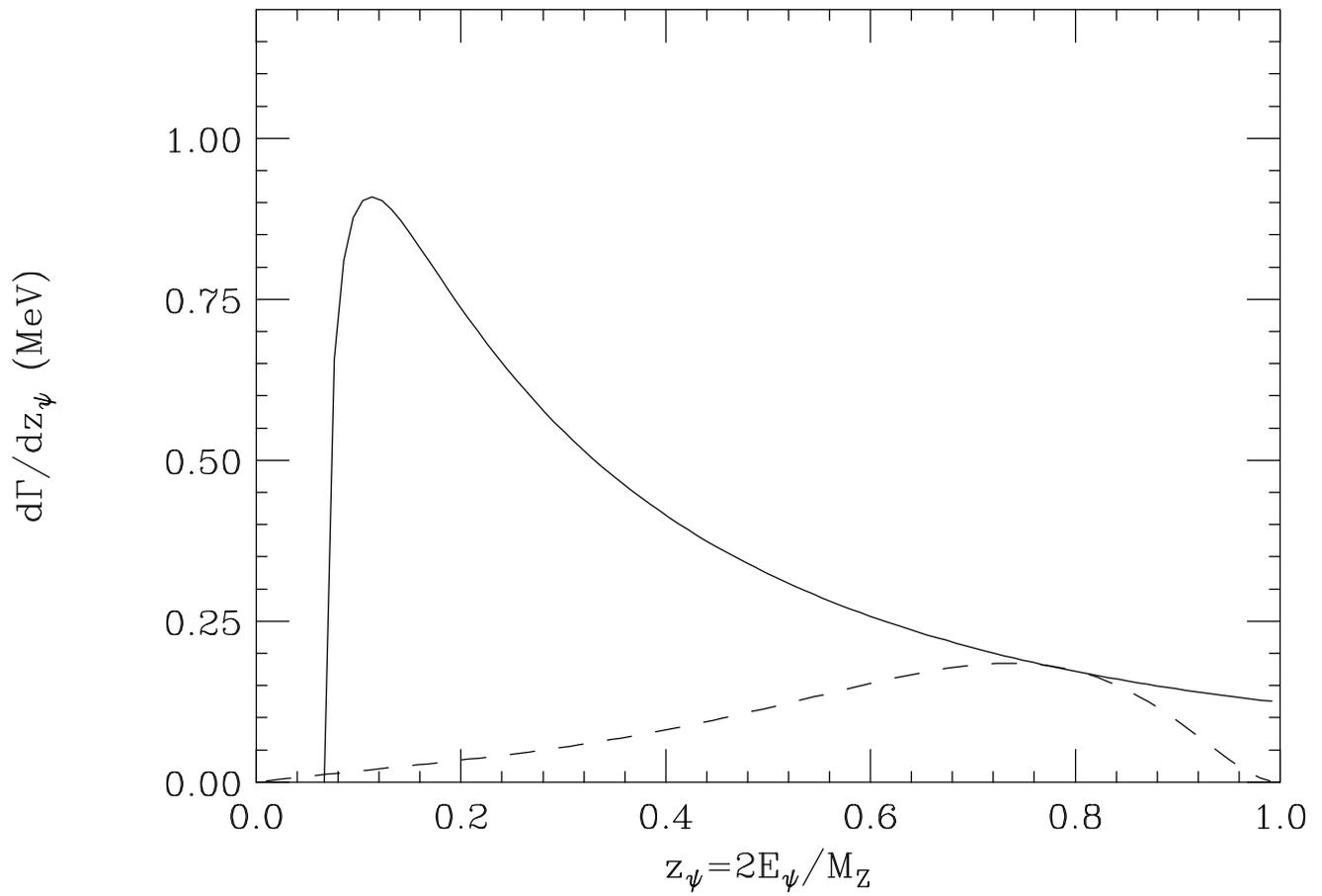
{\it Figure 4} $\;$
Predictions for the energy spectrum $d\Gamma/dz$ for prompt $\psi$'s 
from $Z^0$ decay
in the color-singlet model (dashed line) 
and from the leading color-octet process (solid line). 
%
%

\end{document}